\documentclass[pre,tighten,nofootinbib]{revtex4-1}
\usepackage{graphicx}
\usepackage{amsmath}
\usepackage{amssymb}
\usepackage{bm}
\usepackage{color}
\makeatletter
\DeclareRobustCommand\bblash{\btt{\@backslashchar}}%
\makeatother
\newcommand{\eq}{\text{eq}}
\newcommand{\nq}{\text{neq}}

\begin{document}

\title{Mpemba effect in inertial suspensions}
\author{Satoshi Takada}\email[e-mail:]{takada@go.tuat.ac.jp}
\affiliation{Institute of Engineering, Tokyo University of Agriculture and Technology,
2--24--16, Naka-cho, Koganei, Tokyo 184--8588, Japan}
\author{Hisao Hayakawa}\email[e-mail:]{hisao@yukawa.kyoto-u.ac.jp}
\affiliation{Yukawa Institute for Theoretical Physics, Kyoto University, Kyoto 606--8502, Japan}
\author{Andr\'es Santos}\email[e-mail:]{andres@unex.es}
\affiliation{Departamento de F\'{\i}sica and Instituto de Computaci\'on Cient\'{\i}fica Avanzada (ICCAEx),
Universidad de Extremadura, E--06071 Badajoz, Spain}

\begin{abstract}
The Mpemba effect (a counterintuitive thermal relaxation process where an initially hotter system may cool down to the steady state sooner than an initially colder system) is studied in terms of a model of inertial suspensions under shear. 
The relaxation to a common steady state of a suspension initially prepared in a quasi-equilibrium state is compared with that of a suspension initially prepared in a nonequilibrium sheared state. Two classes of Mpemba effect are identified, the normal and the anomalous one.
The former is generic, in the sense that the kinetic temperature starting from a cold nonequilibrium sheared state is overtaken by the one starting from a hot quasi-equilibrium state, due to the absence of initial viscous heating in the latter, resulting in a faster initial cooling.
The anomalous Mpemba effect is opposite to the normal one since, despite the initial slower cooling of the nonequilibrium sheared state, it can eventually overtake an initially colder quasi-equilibrium state.
The theoretical results based on  kinetic theory agree with those obtained from event-driven simulations for inelastic hard spheres.
It is also confirmed the existence of the inverse Mpemba effect, which is a peculiar heating process, in these suspensions.
More particularly, we find the existence of a mixed process in which both heating and cooling can be observed during relaxation.
\end{abstract}
\date{\today}
\maketitle

\section{Introduction}\label{introduction}
The Mpemba effect is known as an exotic process in which a liquid at a given temperature can freeze faster than another liquid at a lower temperature.
Although there exists a long pre-history of this effect, it became well known after its rediscovery by Mpemba and Osborne~\cite{Mpemba69}.
Several explanations of this effect have been proposed, such as supercooling~\cite{Auerbach95}, properties of hydrogen bonds~\cite{Zhang14}, freezing-point depression by solutes~\cite{Katz09}, a difference in the nucleation temperatures of ice nucleation sites between samples~\cite{Brownridge11}, or a condensed molecular system approaching an equilibrium state with the violation of equipartition law \cite{Gijon19}.
Despite these investigations to support the Mpemba effect, there exists still a certain skepticism on its validity~\cite{Burridge16,Burridge20,Elton20}.

It is obvious that there is no possibility of observing the Mpemba effect if we compare the cooling processes of two equilibrium liquids at different temperatures if they are assumed to be at local equilibrium during relaxation.
If the Mpemba process can be observed, it must be related to nonequilibrium effects.
In other words, the Mpemba effect can be regarded as a peculiar type of relaxation in systems far from equilibrium.
In this sense, we can focus on idealistic situations without freezing effects to extract the essence of the Mpemba effect.
Along this line, the existence of Mpemba-like thermal relaxations has been reported in various systems such as carbon nanotube resonators~\cite{Greaney11}, granular gases with constant~\cite{Santos17, Torrente19, Biswas20} or velocity-dependent~\cite{Mompo20} restitution coefficients, clathrate hydrates~\cite{Ahn16}, dilute atomic gases in an optical resonator~\cite{Keller18}, molecular gases under nonlinear drag \cite{Santos20}, molecular binary gas mixtures~\cite{GomezGonzalez20}, spin glasses~\cite{Baity-Jesi19}, as well as in purely theoretical papers based on Markovian~\cite{Lu17, Klich19, Gal20} or non-Markovian  dynamics~\cite{Lapolla20}, the former of which have recently been experimentally tested \cite{Kumar20, Chetrite21}.
It is remarkable that  the existence of an inverse Mpemba effect, i.e., a paradoxical heating effect in which a material starting from a lower temperature can have a higher temperature than that for a material starting from a higher temperature, has also been reported~\cite{Santos17,Lu17}.

In this paper, we study a class of thermal cooling process, which we call Mpemba effect for simplicity throughout our paper, by analyzing a model of sheared inertial suspension~\cite{Koch01}.
The system we consider might be close to the original setup by Mpemba and Osborne \cite{Mpemba69} because they analyzed a system of ice-mix, which is a suspension system~\cite{Mpemba69}.
Therefore, we believe that our analysis can be appropriate to illustrate some of the universal features of the Mpemba effect.
Through our analysis, we will demonstrate the existence of two types of Mpemba effects, which we term as the normal Mpemba effect (NME) and the anomalous Mpemba effect (AME).

\begin{figure}[htbp]
	\includegraphics[width=0.8\linewidth]{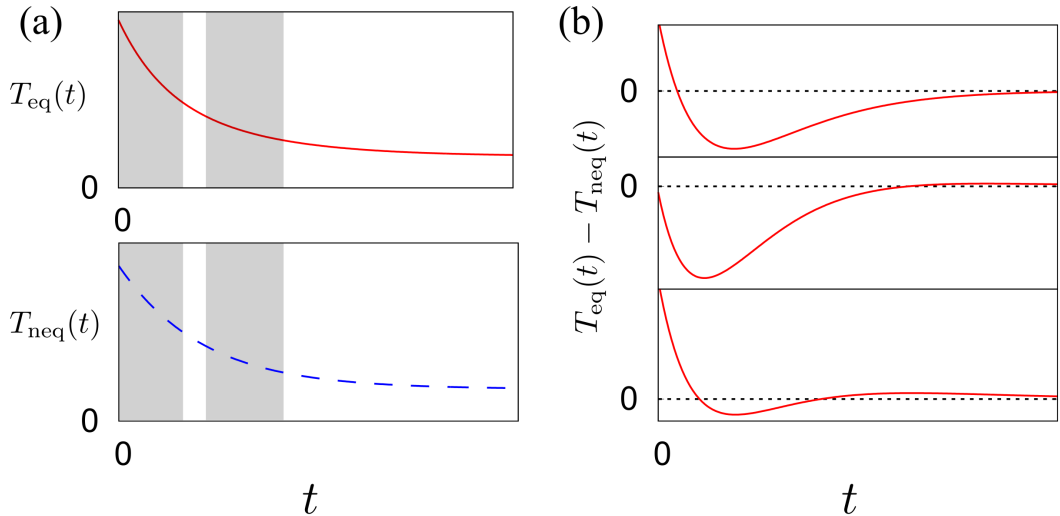}
	\caption{(a) Schematics of the evolutions of the temperatures from the equilibrium $T_{\rm eq}(t)$ (top) and from the nonequilibirum $T_{\rm neq}(t)$ (bottom).
	The two shaded regimes in each figure represent the early and later stages of the relaxation.
	Panel (b) shows three typical possible evolutions of the temperature difference $T_{\rm eq}(t)-T_{\rm neq}(t)$ when the Mpemba effect occurs: (top) NME, (middle) AME, and (bottom) NME+AME.}
	\label{fig:Mpemba}
\end{figure}

It is easy to understand that the NME is generic and can be observed in any sheared suspension as the difference of relaxation rates from initial equilibrium and nonequilibrium conditions.
 Indeed, the time evolution of the temperature ($T$) of suspension liquids under shear may satisfy
\begin{equation}\label{eq_of_T}
	c_V\dot{T}= 
	-\frac{\dot{\gamma}}{n}P_{xy}+2c_V\zeta (T_{\rm env}-T) ,
\end{equation}
where $c_V$, $\dot{\gamma}$, $P_{xy}$, $\zeta$, and $T_{\rm env}$ are the specific heat at constant volume, the shear rate, the shear stress, the drag coefficient, and the environmental temperature, respectively.
Let us consider an equilibrium initial state [where $P_{xy}^\eq(0)=0$] and a nonequilibrium (sheared) initial state [where $P_{xy}^\nq(0)<0$] at temperatures $T_\eq(0)$ and $T_\nq(0)$, respectively, both temperatures being sufficiently higher than the environmental temperature $T_{\rm env}$, so that both initial rates of change, $\dot{T}_\eq(0)$ and $\dot{T}_\nq(0)$, are negative.
Because the first term on the right-hand side of Eq.\ \eqref{eq_of_T} (the viscous heating term) is zero for a system at equilibrium, while it is positive at nonequilibrium, the relaxation rate $|\dot{T}(0)|$ of temperature from an equilibrium initial condition is always larger than that from a nonequilibrium initial condition at the same initial temperature, i.e., $\dot{T}_\eq(0)<\dot{T}_\nq(0)<0$ with $T_\eq(0)=T_\nq(0)$ (see also the first shaded region in Fig.\ \ref{fig:Mpemba}(a) for schematics). 
Therefore, if $T_\eq(0)$ is slightly higher than $T_\nq(0)$, it is even more evident that $\dot{T}_\eq(0)<\dot{T}_\nq(0)<0$, so that during the relaxation under a common shear rate $\dot{\gamma}$ the temperature difference $T_\eq(t)-T_\nq(t)$ initially decreases, thus allowing for the relaxation curve of the equilibrium initial state to catch up that of the nonequilibrium initial state (see the top panel in Fig.\ \ref{fig:Mpemba}(b)).
This is the simple origin of the NME.
With the aid of a parallel argument, if $T_\eq(0)$ is slightly lower than $T_\nq(0)$, then one still has $\dot{T}_\eq(0)<\dot{T}_\nq(0)<0$, so that the initial slope of the equilibrium initial state is more negative than that of the nonequilibrium initial state, i.e., the temperature difference $T_\nq(t)-T_\eq(t)$ increases during the early stage of evolution. In such a case, the relaxation curve $T_\nq(t)$ is arguably expected not to catch up ever the relaxation curve $T_\eq(t)$. 
On the other hand, a nontrivial AME is present if, despite its early increase and as a nonlinear consequence of the relaxation process (see the second shaded region in Fig.\ \ref{fig:Mpemba}(a)), the temperature difference $T_\nq(t)-T_\eq(t)$ eventually vanishes at a certain time (see the middle panel in Fig.\ \ref{fig:Mpemba}(b)).
Interestingly, if the AME is possible with $T_\eq(0)\lesssim T_\nq(0)$, then a temperature crossing must still exist if $T_\eq(0)$ is slightly higher than $T_\nq(0)$, so that the NME will be followed by the second AME crossing.
We will refer to this phenomenon as NME+AME (see the bottom panel in Fig.\ \ref{fig:Mpemba}(b)).
The Mpemba effect previously observed in granular and normal fluids \cite{Santos17,Torrente19,Biswas20,Mompo20,Santos20,GomezGonzalez20} belongs to the NME class. However, to the best of our knowledge, neither AME nor NME+AME has been reported before.
Figure \ref{fig:Mpemba} summarizes schematics of the temperature and the temperature-difference evolutions when those different classes of the Mpemba effect appear. 
Note that the NME+AME is a combination of a transient NME with an asymptotic AME in the sense that $T_{\rm eq}(t)>T_{\rm neq}(t)$ both initially and for asymptotically long times.

We also study the inverse Mpemba effect in this paper.
Similar to the cooling process, the essence of the normal inverse Mpemba effect (NIME) can be understood by Eq.~\eqref{eq_of_T}.
Indeed, the temperature starting from a nonequilibrium initial condition can become higher than the one starting from an equilibrium initial condition, if $T_\nq(0)<T_\eq(0)<T_{\rm env}$, because of the existence of the initial viscous heating effect only in the system starting from the nonequilibrium sheared condition.
An anomalous inverse Mpemba effect (AIME) in heating systems can also exist in analogy to the AME in cooling processes.
Moreover, we confirm the existence of the mixed Mpemba effect (MME), in which temperature inversion takes place during heating and cooling processes in both systems.

The organization of this paper is as follows.
In Sec.\ \ref{sec:Enskogbis}, we introduce the Langevin equation of inertial suspensions and the corresponding Enskog kinetic theory \cite{Garzo}.
The outline of the analysis of the unsteady kinetic theory is also presented in Sec.\ \ref{sec:Enskogbis}, which is parallel to that for steady states \cite{Hayakawa17,Takada20a}.
Section \ref{sec:simulation} is devoted to the study of the NME and the AME through the analysis of the inertial suspension from both kinetic theory and event-driven Langevin simulations for hard spheres (EDLSHS)~\cite{Scala12, DST16, Hayakawa17, Takada20a}.
In Sec.\ \ref{Sec:Inverse_Mixed},  we present results on NIME and AIME for heating processes observed in inertial suspensions.
In that section, we also include the analysis of MME, in which both cooling and heating processes coexist during the time evolution.
Section \ref{sec:discussion} is devoted to discussion and conclusion.
In Appendix \ref{sec:derivation}, we derive moment equations describing the time evolution of the system.
In Appendix \ref{sec:collisionless}, we analyze a collisionless case corresponding to an idealistic model expressing the hydrodynamic lubrication effect to prevent particles from collisions.
In Appendix \ref{sec:Langevin}, we present the relationship between the input parameters of the Langevin equation and the measured temperatures at $t=0$.
In Appendix \ref{sec:Mpemba_time}, we discuss the crossing times on the phase boundaries.
In Appendix \ref{sec:domain}, we illustrate characteristic domain structures near and far from phase boundaries with the aid of an order parameter.
Appendix \ref{sec:eta} is devoted to the discussion on the nontrivial relaxation processes for the viscosity, similar to the Mpemba effect for temperature.

\section{Model: Langevin equation and Enskog kinetic equation for suspensions under simple shear flow}
\label{sec:Enskogbis}

Let us consider a collection of monodisperse smooth hard spheres of diameter $\sigma$, mass $m$, and restitution coefficient $e$ (satisfying $0< e \le 1$) immersed in a three-dimensional fluid.
We assume that the suspended particles are distributed in the background fluid under the influence of a simple shear flow.
This state is macroscopically characterized by a constant number density $n$, a uniform kinetic temperature $T$, and a macroscopic velocity field $\bm{u}=(u_x,\bm{u}_\perp)$ with a constant shear rate $\dot{\gamma}$, namely
\begin{equation}
\label{plane_shear}
	u_x=\dot{\gamma} y, \quad \bm{u}_\perp=\bm{0},
\end{equation}
where $x$ is the shearing direction and $y$ is the direction of change of the sheared velocity.
Here, we assume that the shear flow is symmetrical with respect to $y=0$ and that the boundary effects are not important.
Let us introduce the peculiar momentum of $i$th particle as $\bm{p}_i\equiv m(\bm{v}_{i}-\dot{\gamma} y_i \bm{e}_x)$, where $\bm{v}_i$ is the velocity of $i$th particle and $\bm{e}_x$ is the unit vector parallel to the $x$ direction.
A reliable model for describing suspensions is the Langevin equation:
\begin{equation}
	\label{Langevin_eq}
	\frac{d{\bm{p}}_i}{dt}=-\zeta \bm{p}_i + \bm{F}_i^{({\rm imp})}+ m\bm{\xi}_i,
\end{equation}
where we have assumed that the particles are suspended in the fluid flow for low Reynolds number.
We have also introduced the impulsive force $\bm{F}_i^{(\rm imp)}$ to express collisions between grains, while the noise $\bm{\xi}_i(t)=\xi_{i,\alpha}(t)\bm{e}_\alpha$ has the average properties
\begin{equation}
	\label{noise}
	\langle \bm{\xi}_i(t)\rangle=0, \quad
	\langle \xi_{i,\alpha}(t)\xi_{j,\beta}(t')\rangle
	= \frac{2\zeta T_{\rm env}}{m} \delta_{ij}\delta_{\alpha\beta}\delta(t-t').
\end{equation}
Here, as in Eq.\ \eqref{eq_of_T}, the parameters $\zeta$ and $T_{\rm env}$ characterize the drag from the background fluid and the environmental temperature (in units of energy), respectively.
In reality, the drag coefficient $\zeta$ depends on the moving speed if the latter is high and suspended particles are not small, and should be a resistance matrix as a result of the hydrodynamic interactions between particles, even for slowly moving small suspensions, which strongly depend on their configuration.
This simple model might be applicable to the description of inertial suspensions in which the mean diameter of suspended particles is approximately ranged from 1\,$\mu$m to 70\,$\mu$m~\cite{Koch01}.
For hard-core liquids, it is well known that  $\zeta \propto \eta_0 \propto \sqrt{T_{\rm env}}$, where $\eta_0$ is the viscosity of the solvent or the fluid phase.
Note that the density dependence of $\zeta$ was considered in Refs.~\cite{Hayakawa17, Takada20a}, but the results are qualitatively unchanged from those for a constant $\zeta$.
If we ignore the density dependence of $\zeta$ and the polydispersity of grain sizes,
the Langevin model~\eqref{Langevin_eq} is equivalent to that used by Kawasaki {et al}.~\cite{Kawasaki14}.
For simplicity, we ignore the density dependence of $\zeta$ and the effect of gravity throughout this paper because we have already confirmed that such a dependence is not important \cite{Hayakawa17, Takada20a}.
The latter condition may not be easily achieved for suspensions, but many aerosol particles approximately satisfy it because of their slow sedimentation rates.
Thus, the inertial suspension can be regarded as an idealistic model of aerosol particles.\footnote{Weak attractive interactions between aerosol particles exist, though such an effect is not crucial, as shown in Ref.~\cite{DST16}.}
To solve the Langevin equation \eqref{Langevin_eq} by computer simulations, we adopt EDLSHS, whose outline is summarized in Ref.\ \cite{Scala12} (see also Refs.\ \cite{Hayakawa17, Takada20a}).


Let us rewrite the Langevin equation of the suspension under simple shear flow via the kinetic equation for the one-body  distribution function $f(\bm{r},\bm{v},t)$.
For numerical calculation, the simple shear flow state is generated by Lees--Edwards boundary condition~\cite{LE72}, which is a periodic boundary condition in the local Lagrangian frame characterized by the peculiar velocity $\bm{V}=(v_x-\dot{\gamma} y)\bm{e}_x+\bm{v}_\perp$.
If we assume that the system is uniform in the Lagrangian frame, the velocity distribution function satisfies
\begin{equation}
\label{vic1}
	f(\bm{r},\bm{v},t)=f(\bm{V},t),
\end{equation}
and the Enskog equation for the granular suspension is~\cite{Hayakawa17,Takada20a,Hayakawa03,Chamorro15,DST16}
\begin{equation}
	\label{Enskog}
	\left(\frac{\partial}{\partial t}-\dot{\gamma} V_{y}\frac{\partial}{\partial V_{x}}\right)f(\bm{V},t)
	=\zeta\frac{\partial}{\partial \bm{V}} \cdot
	\left[\left( \bm{V}+ \frac{T_{\rm env}}{m} \frac{\partial}{\partial \bm{V}} \right) f(\bm{V},t) \right]
	+ J_\text{E}[\bm{V}|f,f],
\end{equation}
where the Enskog collision operator $J_\text{E}[\bm{V}|f,f]$ is given by~\cite{Hayakawa17,Takada20a}
\begin{equation}
	\label{J(V|f)}
	J_{\text{E}}\left[\bm{V}_{1}|f,f\right]
	=\sigma^2g_0 \int d\bm{V}_{2}\int d\widehat{\boldsymbol{\sigma}}\,
	\Theta (\widehat{{\boldsymbol {\sigma}}} \cdot \bm{V}_{12})
	(\widehat{\boldsymbol {\sigma }}\cdot \bm{V}_{12})
	\left[\frac{f(\bm{V}_1^{\prime\prime},t)f(\bm{V}_2^{\prime\prime}+\dot{\gamma}\sigma \widehat{\sigma}_y \bm{e}_x,t)}{e^2}
	-f(\bm{V}_1,t)f(\bm{V}_2-\dot{\gamma}\sigma \widehat{\sigma}_y \bm{e}_x,t)\right].
\end{equation}
Here, $g_0$ is the radial distribution at contact for hard spheres, whose (approximate) explicit expression is given by~\cite{CS}
\begin{equation}\label{radial_fn}
	g_0(|\bm{r}|=\sigma,\varphi)=\frac{1-\varphi/2}{(1-\varphi)^3},
\end{equation}
with the volume fraction $\varphi=(\pi/6) n\sigma^3$ satisfying $\varphi< 0.49$.
In Eq.\ \eqref{J(V|f)}, we have introduced the Heaviside step function defined as $\Theta(x)=1$ for $x\ge 0$ and $\Theta(x)=0$ otherwise,
the relative velocity at contact $\bm{V}_{12}=\bm{V}_1-\bm{V}_2$,
 and the unit vector $\widehat{\bm{\sigma}}=(\bm{r}_2-\bm{r}_1)/\sigma$ at contact.
In addition, the double primes in Eq.\ \eqref{J(V|f)} denote the pre-collisional velocities $\left\{\bm{V}_1^{\prime\prime}, \bm{V}_2^{\prime\prime}\right\}$, which satisfy the following collision rule:
\begin{equation}
	\label{collision_rule}
	\bm{V}_1^{\prime\prime}=\bm{V}_1-\frac{1+e}{2e}(\bm{V}_{12}\cdot\widehat{\bm{\sigma}})\widehat{\bm{\sigma}}, \quad
	\bm{V}_2^{\prime\prime}=\bm{V}_2+\frac{1+e}{2e}(\bm{V}_{12}\cdot\widehat{\bm{\sigma}})\widehat{\bm{\sigma}},
\end{equation}
with $\left\{\bm{V}_1, \bm{V}_2\right\}$ being the post-collisional velocities of particles 1 and 2.
In this paper, we do not consider the effects of tangential friction and rotation induced by each binary collision.

The most important quantity to characterize the shear flow is the stress tensor $\mathsf{P}$. It has kinetic and collisional transfer contributions, i.e., $\mathsf{P}=\mathsf{P}^k+\mathsf{P}^c$.
Here, the kinetic stress $\mathsf{P}^k$ is given by
\begin{equation}
	\label{pressure_tensor:kinetic}
	P^k_{\alpha\beta}=m \int d\bm{V} V_\alpha V_\beta f(\bm{V}),
\end{equation}
while its collisional contribution $\mathsf{P}^c$ to the stress is given by~\cite{Garzo,Hayakawa17,Takada20a,Santos98, Montanero99}
\begin{equation}
	\label{pressure:collisional}
	P^c_{\alpha\beta}
	=\frac{1+e}{4} m\sigma^3 g_0 \int d\bm{V}_1 \int d\bm{V}_2\int d\widehat{\bm{\sigma}}
	\Theta(\bm{V}_{12}\cdot\widehat{\bm{\sigma}})
	(\bm{V}_{12}\cdot\widehat{\bm{\sigma}})^2\widehat{\sigma}_\alpha\widehat{\sigma}_\beta
	f\left(\bm{V}_1\right)
	f\left(\bm{V}_2-\dot{\gamma} \sigma\widehat{\sigma}_y \bm{e}_x\right).
\end{equation}
The hydrostatic pressure $P$ is defined as $P\equiv P_{\alpha\alpha}/3$,
where we adopt Einstein's rule for the summation, i.e., $P_{\alpha\alpha}=\sum_{\alpha=1}^3 P_{\alpha\alpha}$.
The kinetic pressure satisfies the equation of state of ideal gases, namely,  $P^k\equiv P^k_{\alpha\alpha}/3 =n T$, where
\begin{equation}
	\label{vic3}
	n=\int d\bm{V} f(\bm{V}),\quad T =\frac{m}{3n}\int d\bm{V} \bm{V}^2f(\bm{V})
\end{equation}
are the number density and the kinetic temperature, respectively.
Throughout this paper, we assume that the kinetic temperature is measurable and is used to detect the Mpemba effect.

It should be noted that the model of inertial suspension with $T_{\rm env}=0$ was introduced in Ref.~\cite{Tsao95} for dilute suspensions and in Ref.~\cite{Sangani96} for moderately dense suspensions.
See also Ref.~\cite{Saha17} for dilute inertial suspensions with $T_{\rm env}=0$.
On the other hand, the model with $T_{\rm env}=0$ has several defects because (i) suspensions are not stable against clustering if there are no thermal agitations, (ii) the viscosity and the drag become zero in the zero-temperature limit, and (iii) thermal equilibrium states cannot be recovered in the unsheared situation.
The series of our recent papers~\cite{Hayakawa17, Takada20a, DST16, Sugimoto20} can be regarded as an up-to-date analysis for \emph{steady} states of inertial suspensions at finite densities.
In particular, Ref.~\cite{Takada20a} demonstrated that the Enskog kinetic theory gives very precise descriptions of steady states for moderately dense suspensions.
Therefore, we can apply the kinetic theory in \emph{unsteady} states to describe relaxation processes in inertial suspensions.
Hereafter, we solve the time evolution of the system obtained from Eq.\ \eqref{Enskog} (see the detailed expressions in Appendix \ref{sec:derivation}).

\section{Theoretical and numerical results of the Mpemba effect in cooling processes}\label{sec:simulation}

In this section, we show that the Mpemba effect takes place through our EDLSHS of Eq.\ \eqref{Langevin_eq} with Eq.\ \eqref{noise}, as well as through the moment equations from the Enskog kinetic theory in Grad's approximation.
We also demonstrate that the time evolutions obtained from the theory reproduce well those from the EDLSHS.

\begin{figure}[htbp]
	\includegraphics[width=0.7\linewidth]{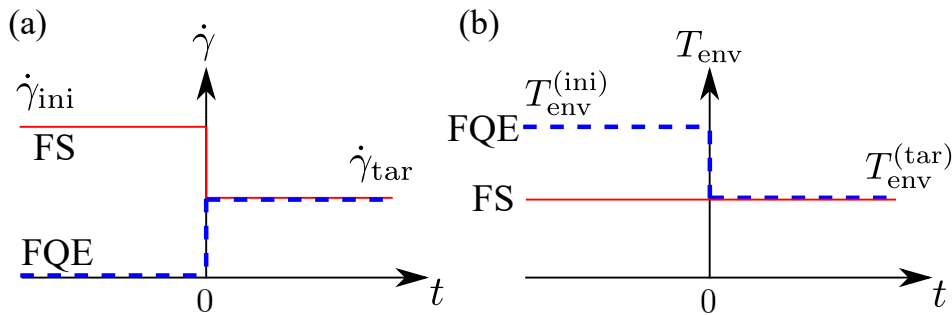}
	\caption{Schematic illustrations of our protocol for FQE (dashed lines) and FS (solid line): the shear rate and the environmental temperature are changed at $t=0$ (a) from $0$ to $\dot{\gamma}_{\rm tar}$ and (b) from $T_{\rm env}^{(\rm ini)}$ to $T_{\rm env}^{(\rm tar)}$, respectively.
	(a) the shear rate is changed at $t=0$ from $\dot{\gamma}_{\rm ini}$ to $\dot{\gamma}_{\rm tar}$ while (b) the environmental temperature is kept at $T_{\rm env}^{(\rm tar)}$.
	Here,  the case for $\dot{\gamma}_{\rm ini}>\dot{\gamma}_{\rm tar}$ and $T_{\rm env}^{(\rm ini)}>T_{\rm env}^{(\rm tar)}$ is presented.}
	\label{fig:protocolB}
\end{figure}

Let us consider the following protocol.
In general, we can examine a situation in which both the shear rate and environmental temperature for $t<0$, i.e., $\dot{\gamma}_{\rm ini}$ and $T_{\rm env}^{(\rm ini)}$, might be different from the ``target'' values  for $t>0$, i.e.,  $\dot{\gamma}_{\rm tar}$ and $T_{\rm env}^{(\rm tar)}$.
Two specific choices will be considered here.
First, we denote with the label FQE  a system starting from a \emph{quasi-equilibrium} steady initial state, in which the shear rate and the environmental temperature are changed at $t=0$ from $0$ to $\dot{\gamma}_{\rm tar}$ and from $T_{\rm env}^{(\rm ini)}$ to $T_{\rm env}^{(\rm tar)}$, respectively.
Analogously, the label FS denotes a system starting from a \emph{sheared} steady initial state such that the shear rate is changed at $t=0$ from $\dot{\gamma}_{\rm ini}$ to $\dot{\gamma}_{\rm tar}$, while the environmental temperature is unchanged and made equal to $T_{\rm env}^{(\rm tar)}$.
Figure \ref{fig:protocolB} provides schematic illustrations of this protocol for both systems.

The initial condition for the system FQE is obtained by the time evolution equations \eqref{Langevin_eq} associated with the noise condition \eqref{noise} (in the case of simulations) and \eqref{Enskog} (in the case of kinetic theory) by setting $\dot{\gamma}=0$ and $T_{\rm env}=T_{\rm env}^{(\rm ini)}$, and allowing the system to reach a quasi-equilibrium steady state by the balance between the noise and the dissipation induced by each collision.
In turn, the initial condition for the FS system  is obtained by setting $\dot{\gamma}=\dot{\gamma}_{\rm ini}$ and $T_{\rm env}=T_{\rm env}^{(\rm tar)}$, and waiting until  a sheared steady state is reached.
It should be noted that there is no shear stress for unsheared quasi-equilibrium systems,\footnote{This is easy to be understood.
First, it is obvious that the kinetic stress introduced in Eq.\ \eqref{pressure_tensor:kinetic} satisfies $P^k_{\alpha\beta}=0$ for $\alpha\ne \beta$ if we assume that the velocity distribution function is invariant under the change $V_\alpha\to -V_\alpha$.
Next, the off-diagonal collisional stress $P^c_{\alpha\beta}$ introduced in Eq.\ \eqref{pressure:collisional} vanishes because $P^c_{\alpha\beta}=-P^c_{\alpha\beta}$ under the changes of $1\leftrightarrow 2$ and $\widehat{\bm{\sigma}}\leftrightarrow -\widehat{\bm{\sigma}}$.
Thus, $P_{\alpha\beta}=0$ for $\alpha\ne \beta$ if the velocity distribution is isotropic.}
while $P_{xy}(0)<0$ in the FS system.
In this paper, we generally fix the value of the target environmental temperature as $T_{\rm env}^{(\rm tar)*}=1.0$, where $T_{\rm env}^*\equiv T_{\rm env}/(m\sigma^2\zeta^2)$.

Now, let us introduce the initial temperature ratio $\vartheta$ as
\begin{equation}
	\vartheta\equiv \frac{T_{\rm FQE}(0)}{T_{\rm FS}(0)}=\frac{\theta_{\rm FQE}(0)}{\theta_{\rm FS}(0)},
	\label{eq:vartheta}
\end{equation}
where the dimensionless temperature is defined as $\theta\equiv T/T_{\rm env}$ and we take the environmental temperature for $t>0$, i.e., $T_{\rm env}^{(\rm tar)}$, in both systems.
According to the discussion below Eq.\ \eqref{eq_of_T}, in cooling processes (i.e., $\dot{\gamma}_{\rm ini}>\dot{\gamma}_{\rm tar}$) with $\vartheta>1$, a NME can be present in the transient dynamics before both systems reach a common steady state with $\theta_{\rm tar}\equiv\lim_{\tau\to\infty}\theta(\tau)$, where $\tau\equiv \zeta t$ is the dimensionless time.
Analogously, in heating processes (i.e., $\dot{\gamma}_{\rm ini}<\dot{\gamma}_{\rm tar}$) a NIME is possible if again $\vartheta>1$.
Much less trivial is the possibility of the (cooling) AME or (heating) AIME with $\vartheta<1$. A fully analytical exact treatment is possible in the case of a collisionless model, as worked out in Appendix \ref{sec:collisionless}.
Note that $\vartheta$ is an observable quantity by the measurement of the temperatures at $t=0$, though they are the outcomes of our systems determined by their respective steady states reached for $t<0$.
The relationship between the input parameters $\{T_{\rm env}^{({\rm ini}*)},T_{\rm env}^{({\rm tar}*)},\dot{\gamma}_{\rm ini}^*\}$ (see Fig.\ \ref{fig:protocolB} for $t<0$) and the outcome $\vartheta$ is discussed in Appendix \ref{sec:Langevin}, where we have introduced the dimensionless shear rate $\dot\gamma^*\equiv \dot\gamma/\zeta$.

Given the values of the packing fraction $\varphi$, the restitution coefficient $e$, the target dimensionless environmental temperature $T_{\rm env}^{\rm (tar)*}$, and the target dimensionless shear rate $\dot{\gamma}^*_{\rm tar}$, the time evolutions starting from each pair of initial conditions $\{\text{FS},\text{FQE}\}$ are characterized by only two parameters, namely $\dot{\gamma}^*_{\rm ini}$ and $T_{\rm env}^{({\rm ini})*}$.
In some situations, however, we will use $\dot\gamma^*_{\rm ini}$ and $\vartheta$ as control parameters.
In our simulations, we examine $N=100$ different (microscopic) initial configurations for each initial condition to get average values and error bars.

In the protocol, there are four possibilities depending on the relative values  of (i) $\dot{\gamma}_{\rm ini}^*$ and $\dot{\gamma}_{\rm tar}^*$, and (ii) $T_{\rm env}^{(\rm ini)*}$ and $T_{\rm env}^{(\rm tar)*}$.
We note that the heating case for $\dot{\gamma}_{\rm ini}^*<\dot{\gamma}_{\rm tar}^*$ and $T_{\rm env}^{(\rm ini)}<T_{\rm env}^{(\rm tar)}$ will be discussed as the inverse Mpemba effect in Sec.\ \ref{Sec:Inverse_Mixed}.
We will also discuss there the mixed Mpemba effect in which both cooling and heating processes can be observed during the time evolution.

\begin{figure}[htbp]
	\includegraphics[width=0.9\linewidth]{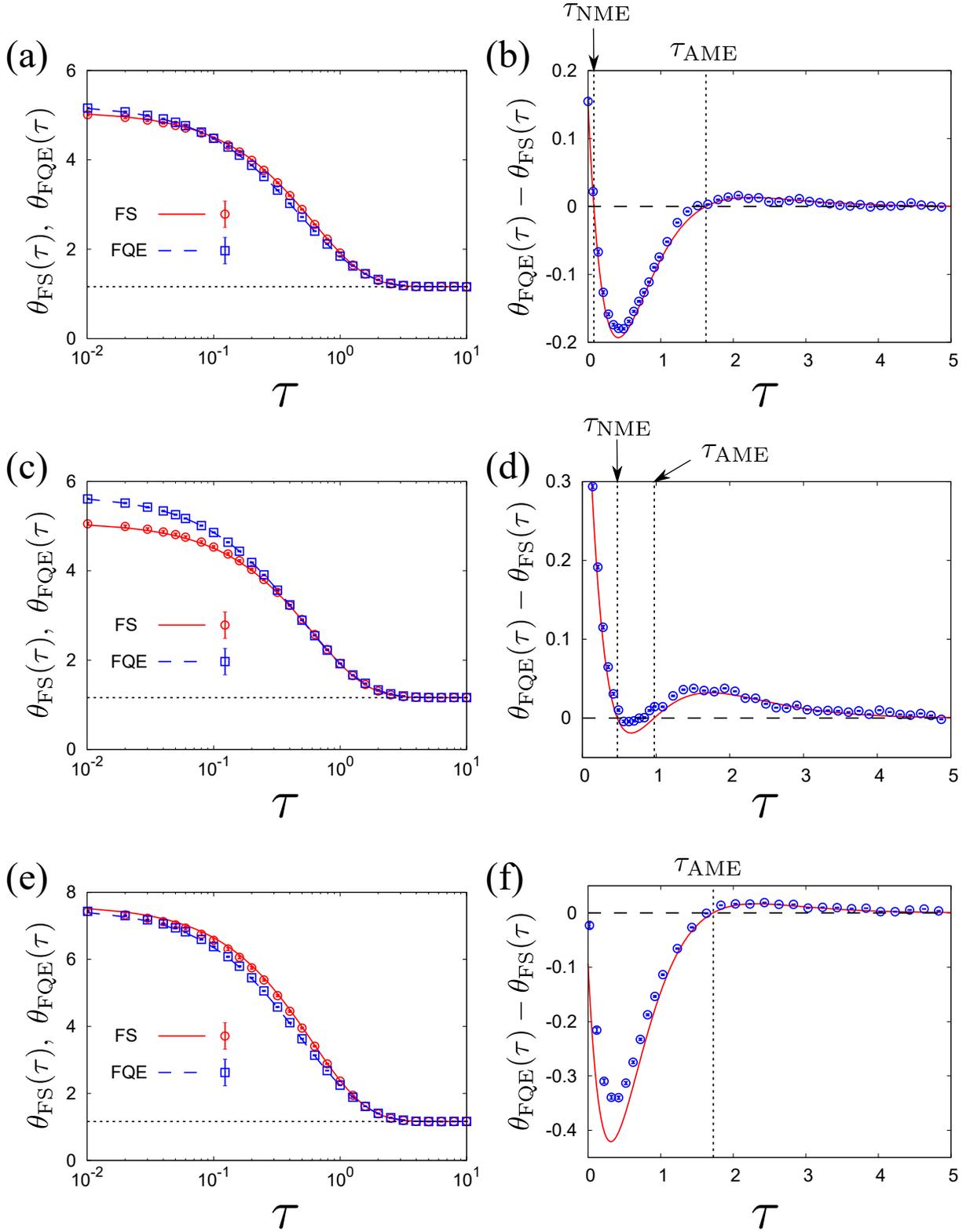}
	\caption{
	Time evolutions of (a, c, e) the temperatures $\theta_{\rm FS}(\tau)$ and $\theta_{\rm FQE}(\tau)$ and (b, d, f) the temperature difference $\theta_{\rm FQE}(\tau)-\theta_{\rm FS}(\tau)$ for $e=0.9$, $\varphi=0.01$, $T_{\rm env}^{(\rm tar)*}=1.0$, and $\dot{\gamma}_{\rm tar}^*=1.0$. 
	Panels (a, b), (c, d), and (e, f) correspond to $(\dot{\gamma}_{\rm ini}^*, T_{\rm env}^{(\rm ini)*})=(4.0, 5.29)$, $(4.0, 5.76)$, and $(4.5, 7.62)$, respectively.
	The dotted lines in panels (a), (c), and (e) represent the target temperature $\theta_{\rm tar}=1.16$.
	The vertical dotted lines in panel (b), (d), and (f) indicate $\tau_{\rm NME}$ and $\tau_{\rm AME}$ defined by Eqs.\ \eqref{eq:tau_NME_AME_def1} and \eqref{eq:tau_NME_AME_def2}, respectively.
	The symbols (with error bars) are obtained from our simulations and the lines are kinetic-theory predictions.}
	\label{fig:evol_xi}
\end{figure}

Figure \ref{fig:evol_xi} shows typical time evolutions of $\theta_{\rm FQE}(\tau)$ and $\theta_{\rm FS}(\tau)$ for $\vartheta>1$ (panels (a)--(d)) and $\vartheta<1$ (panels (e) and (f)), respectively, with fixed $\varphi=0.01$.
The theory reproduces well the time evolution of the temperature obtained from our simulations for both $\vartheta>1$ and $\vartheta<1$.
For $\vartheta>1$, after the initial stage where $\theta_{\rm FQE}(\tau)-\theta_{\rm FS}(\tau)>0$, NME takes place, so that  $\theta_{\rm FQE}(\tau)-\theta_{\rm FS}(\tau)<0$ in an intermediate time window $\tau_{\rm NME}<\tau<\tau_{\rm AME}$.
Then, an AME occurs for $\tau>\tau_{\rm AME}$ in this set of parameters.
Here, we have introduced $\tau_{\rm NME}$, which satisfies
\begin{equation}
	\theta_{\rm FQE}(\tau_{\rm NME}) - \theta_{\rm FS}(\tau_{\rm NME})=0,\quad
	\theta_{\rm FQE}(\tau) - \theta_{\rm FS}(\tau)>0 \quad (\tau<\tau_{\rm NME}),
	\label{eq:tau_NME_AME_def1}
\end{equation}
and $\tau_{\rm AME}$ which satisfies
\begin{equation}
	\theta_{\rm FQE}(\tau_{\rm AME}) - \theta_{\rm FS}(\tau_{\rm AME})=0,\quad
	\theta_{\rm FQE}(\tau) - \theta_{\rm FS}(\tau)<0 \quad
	\begin{cases}
		(\tau_{\rm NME}<\tau<\tau_{\rm AME}:\ ``{\rm NME+AME}")\\
		(\tau<\tau_{\rm AME}:\ ``{\rm AME}")
	\end{cases}.
	\label{eq:tau_NME_AME_def2}
\end{equation}
We call ``NME+AME'' the parameter region where this double crossing takes place.
It is noted that $\tau_{\rm NME}$ ($\tau_{\rm AME}$) is the first (second) crossing time in ``NME+AME'' region (see Figs.\ \ref{fig:evol_xi}(b) and (d)).
Depending on the initial environmental temperature $T_{\rm env}^{({\rm ini})*}$, the magnitude of the AME can be much smaller than that of the NME, as in Fig.~\ref{fig:evol_xi}(b), or almost the same as that of the NME, as in Fig.~\ref{fig:evol_xi}(d).
For the latter case, the time domain of the AME is longer than that of the NME.
For $\vartheta<1$, on the other hand, $\theta_{\rm FQE}(\tau)-\theta_{\rm FS}(\tau)<0$ in the early stage, and it takes a negative peak in the intermediate stage.
Then, $\theta_{\rm FQE}(\tau)$ becomes larger than $\theta_{\rm FS}(\tau)$ after $\tau$ exceeds a crossover value $\tau_{\rm AME}$. 
This is the AME for $\vartheta<1$.
As far as we investigated, the magnitude of AME is smaller than that of the negative peak in the intermediate stage (see Fig.\ \ref{fig:evol_xi}(f)).

\begin{figure}[htbp]
	\includegraphics[width=0.9\linewidth]{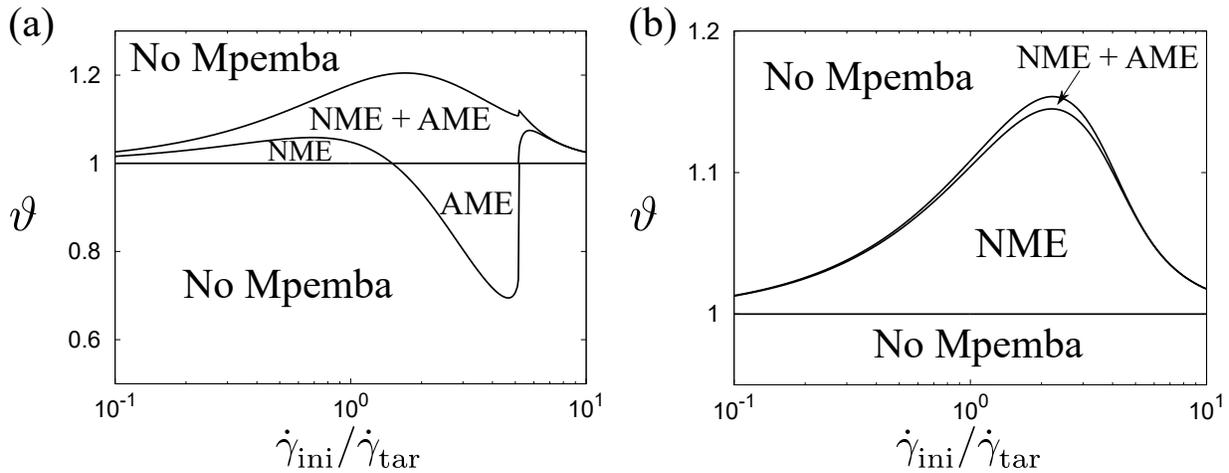}
	\caption{Phase diagrams in the plane $\vartheta$ versus $\dot{\gamma}_{\rm ini}/\dot\gamma_{\rm tar}$ of the Mpemba effect for (a) $\varphi=0.01$ and (b) $\varphi=0.10$ with $e=0.9$, $T_{\rm env}^{(\rm tar)*}=1.0$, and $\dot{\gamma}^*_{\rm tar}=1.0$.}
	\label{fig:phase_xi}
\end{figure}

The phase diagrams in the plane $\vartheta$ versus $\dot{\gamma}_{\rm ini}/\dot\gamma_{\rm tar}$ are shown in Fig.\ \ref{fig:phase_xi} for $e=0.9$, $T_{\rm env}^{(\rm tar)*}=1.0$,  $\dot{\gamma}^*_{\rm tar}=1.0$, and (a) $\varphi=0.01$ and (b) $\varphi=0.1$.
There are four distinct regions in Fig.~\ref{fig:phase_xi}(a), ``No Mpmeba,'' ``NME+AME,'' ``NME,'' and ``AME.''
Note that, although cooling (heating) relaxation processes are located to the right (left) of an imaginary  vertical line $\dot{\gamma}_{\rm ini}/\dot{\gamma}_{\rm tar}=1$, for the sake of simplicity, the notation employed in Fig.\ \ref{fig:phase_xi} does not distinguish between direct and inverse Mpemba effects.
We discuss characteristic behavior of crossing times on the phase boundaries among ``NME+AME,'' ``NME,'' and ``AME'' in Appendix \ref{sec:Mpemba_time}.
We also visualize characteristic domain growths near and far from phase boundaries in Appendix \ref{sec:domain}.

\begin{figure}[htbp]
	\includegraphics[width=0.5\linewidth]{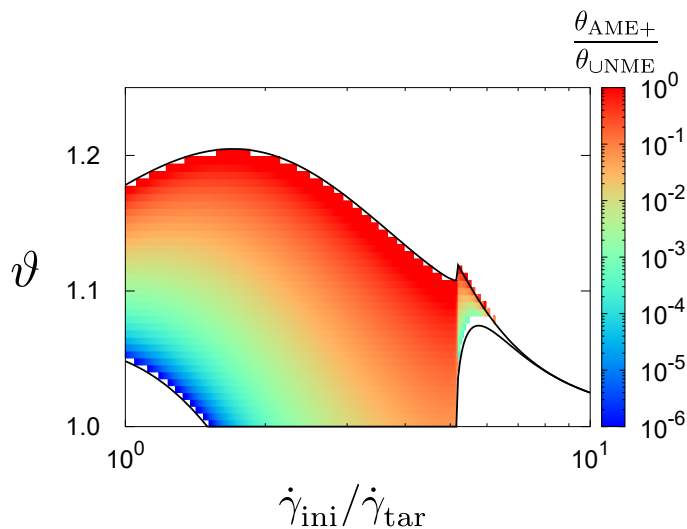}
	\caption{Magnitude plot of the ratio of the amplitudes defined in Eqs.\  \eqref{eq:theta_NME_prime&eq:theta_AME_prime} for the ``NME+AME'' region of Fig.\ \ref{fig:phase_xi}(a) with $\dot{\gamma}_{\rm ini}/\dot{\gamma}_{\rm tar}>1$.}
	\label{fig:phase_xi_0.01}
\end{figure}

Let us define the two amplitudes in the ``NME+AME'' region as
\begin{subequations}
\label{eq:theta_NME_prime&eq:theta_AME_prime}
\begin{align}
	\theta_{\cup{\rm NME}}
	&\equiv \max_{\tau_{\rm NME}<\tau<\tau_{\rm AME}}
	\left\{\theta_{\rm FS}(\tau)-\theta_{\rm FQE}(\tau) \right\},	\label{eq:theta_NME}\\
	\theta_{{\rm AME}+}
	&\equiv \max_{\tau>\tau_{\rm AME}}
	\left\{\theta_{\rm FQE}(\tau)-\theta_{\rm FS}(\tau) \right\}. \label{eq:theta_AME}
\end{align}
\end{subequations}
Figure \ref{fig:phase_xi_0.01} shows a magnitude plot of the ratio $\theta_{{\rm AME}+}/\theta_{\cup{\rm NME}}$ in the subregion with $\dot{\gamma}_{\rm ini}/\dot{\gamma}_{\rm tar}>1$ of the ``NME+AME'' region of Fig.\ \ref{fig:phase_xi}(a).
We observe that the ratio $\theta_{{\rm AME}+}/\theta_{\cup{\rm NME}}$ is small near the boundary between the ``NME'' and ``NME+AME'' regions, but it increases to values $\theta_{{\rm AME}+}/\theta_{\cup{\rm NME}}\simeq 1$ near the boundary between the ``No Mpemba'' and ``NME+AME'' regions.

\begin{figure}[htbp]
	\includegraphics[width=0.5\linewidth]{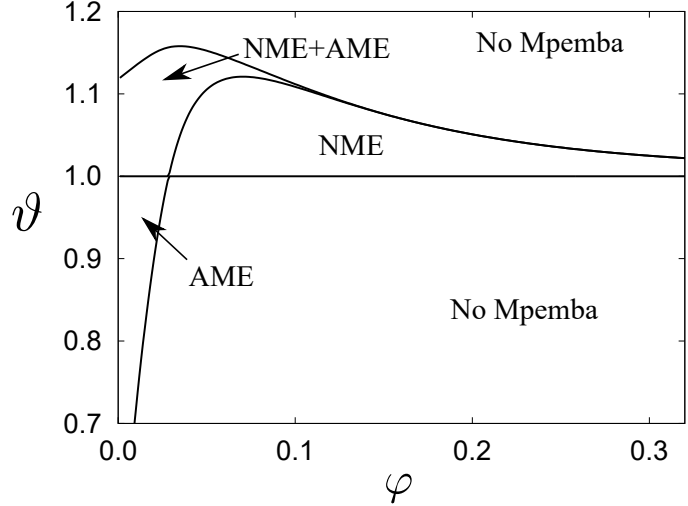}
	\caption{Phase diagram of the Mpemba effect on the plane $\vartheta$ versus $\varphi$ for $e=0.9$, $T_{\rm env}^{(\rm tar)*}=1.0$,  $\dot{\gamma}^*_{\rm tar}=1.0$, and $\dot{\gamma}^*_{\rm ini}=4.0$.}
	\label{fig:phase_density}
\end{figure}

These exotic properties are suppressed as the density increases, as already observed in Fig.\ \ref{fig:phase_xi}(b), where the ``AME'' region is extinct and the ``NME+AME'' region  has almost disappeared at $\varphi=0.10$.
This is also understood from Fig.\ \ref{fig:phase_density}, where the phase diagram in the plane $\vartheta$ versus $\varphi$ is shown for $e=0.9$, $T_{\rm env}^{(\rm tar)*}=1.0$,  $\dot{\gamma}^*_{\rm tar}=1.0$, and $\dot{\gamma}^*_{\rm ini}=4.0$. 
For this choice of parameters, the ``AME'' region disappears at $\varphi=2.8\times10^{-2}$.

\begin{figure}[htbp]
	\includegraphics[width=0.5\linewidth]{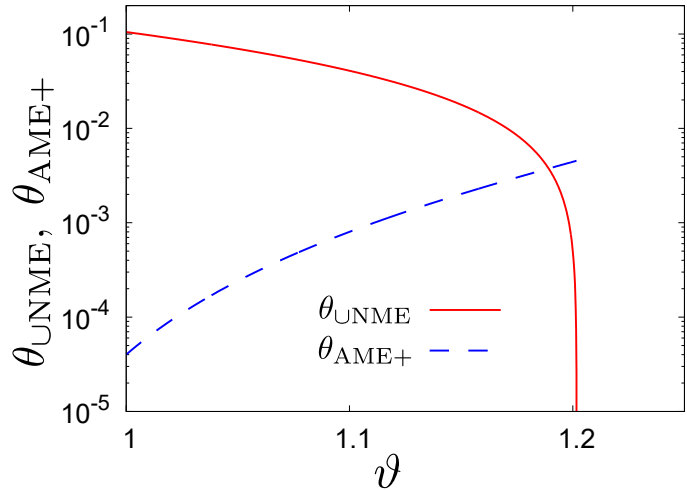}
	\caption{Amplitudes of the temperature differences introduced in Eqs.\ \eqref{eq:theta_NME_prime&eq:theta_AME_prime} as functions of $\vartheta$ for $e=0.9$, $\varphi=0.01$,   $T_{\rm env}^{({\rm tar})*}=1.0$,  $\dot{\gamma}^*_{\rm tar}=1.0$, and $\dot{\gamma}^*_{\rm ini}=2.0$. In agreement with Fig.\ \ref{fig:phase_xi}(a), at the value $\dot{\gamma}^*_{\rm ini}=2.0$ only the ``NME+AME'' region is present with $\vartheta>1$.}
	\label{fig:magnitude}
\end{figure}

Figure \ref{fig:magnitude} represents the amplitudes  $\theta_{\cup{\rm NME}}$ and $\theta_{{\rm AME}+}$ as functions of the ratio $\vartheta$ for $e=0.9$, $\varphi=0.01$,  $T_{\rm env}^{({\rm tar})*}=1.0$,  $\dot{\gamma}^*_{\rm tar}=1.0$, and $\dot{\gamma}^*_{\rm ini}=2.0$.
It is observed that $\theta_{{\rm AME}+} \ll \theta_{\cup{\rm NME}}$ for small $\vartheta-1$, while $\theta_{{\rm AME}+}$ can be larger than $\theta_{\cup{\rm NME}}$ in the vicinity of the boundary between the ``NME+AME'' and ``No Mpemba'' regions.

In order to characterize with a single parameter how the union of the  ``NME'' and ``NME+AME'' regions, on the one hand, and the ``AME'' region, on the other hand, change with the volume fraction, let us introduce the quantities $\Delta \vartheta_{\rm NME}$ and $\Delta \vartheta_{\rm AME}$  as
\begin{equation}\label{Delta_vertheta}
	\displaystyle \Delta \vartheta_{\rm NME}(\varphi) 
	\equiv \max_{\dot{\gamma}^*_{\rm ini}} \left\{\vartheta_{\rm NME}(\varphi,\dot{\gamma}^*_{\rm ini})-1 \right\},
	\quad
	\displaystyle \Delta \vartheta_{\rm AME}(\varphi) 
	\equiv \max_{\dot{\gamma}^*_{\rm ini}} \left\{1-\vartheta_{\rm AME}(\varphi,\dot{\gamma}^*_{\rm ini}) \right\},
\end{equation}
where $\vartheta_{\rm NME}(\varphi,\dot{\gamma}^*_{\rm ini})>1$ and $\vartheta_{\rm AME}(\varphi,\dot{\gamma}^*_{\rm ini})<1$ are values of $\vartheta$ inside the regions ``NME'' (or ``NME+AME'') and ``AME,'' respectively.

\begin{figure}[htbp]
	\includegraphics[width=0.5\linewidth]{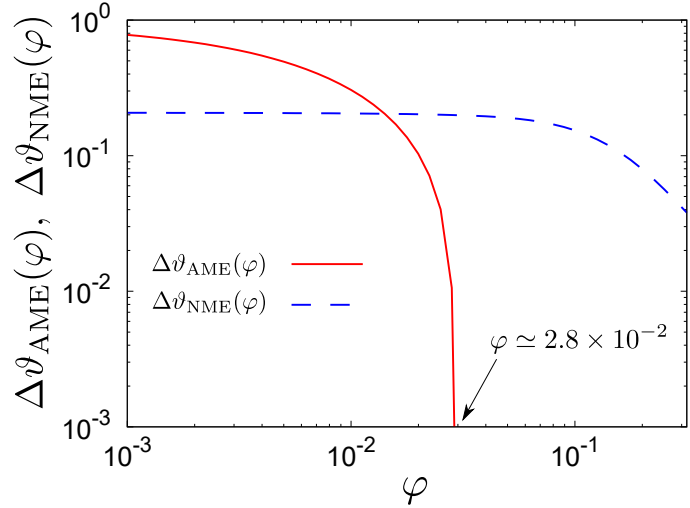}
	\caption{Volume fraction dependence of $\Delta \vartheta_{\rm NME}(\varphi)$ and $\Delta \vartheta_{\rm AME}(\varphi)$, introduced in Eq.~\eqref{Delta_vertheta}, for $e=0.9$, $T_{\rm env}^{({\rm tar})*}=1.0$, and $\dot{\gamma}^*_{\rm tar}=1.0$.}
	\label{fig:order}
\end{figure}

As shown in Fig.\ \ref{fig:order}, $\Delta \vartheta_{\rm NME}(\varphi)$ is weakly dependent on $\varphi$, being finite even for large $\varphi$. However,
$\Delta \vartheta_{\rm AME}(\varphi)$ decays with increasing density and rapidly approaches zero at a threshold volume fraction $\varphi_{\rm AME}\simeq 2.8\times 10^{-2}$
(in the case $e=0.9$,  $T_{\rm env}^{(\rm tar)*}=1.0$, and $\dot{\gamma}^*_{\rm tar}=1.0$). Thus, the region  ``AME''  can be observed only for dilute cases.
Nevertheless, as illustrated by Fig.~\ref{fig:phase_xi}(b), a narrow region ``NME+AME'' is still present for non-dilute systems.

\begin{figure}[htbp]
	\includegraphics[width=\linewidth]{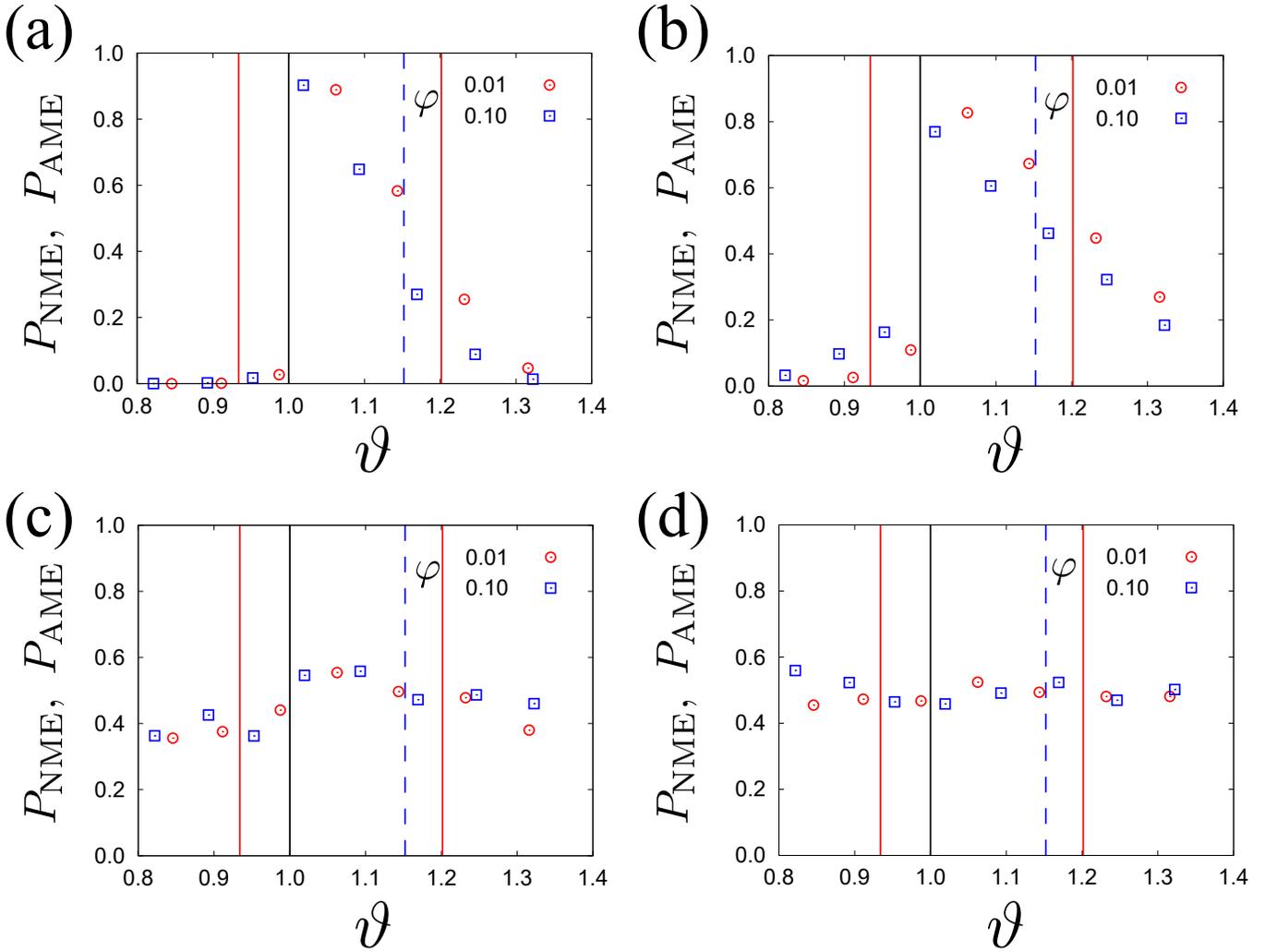}
	\caption{Probabilities of the NME  ($\vartheta>1$) and AME  ($\vartheta<1$) as functions of $\vartheta$ [see Eqs.\ \eqref{eq:P_NME} and \eqref{eq:P_AME}]
at (a) $\tau=0.5$, (b) $\tau=1.0$, (c) $\tau=2.0$, and (d) $\tau=5.0$. The parameters are chosen as $e=0.9$, $T_{\rm env}^{(\rm tar)*}=1.0$, $\dot{\gamma}_{\rm tar}^*=1.0$, $\dot{\gamma}_{\rm ini}^*=2.0$, and two values of the volume fraction ($\varphi=0.01$ and $\varphi=0.10$).	The vertical dotted line signals the boundary value $\vartheta=1$, while the vertical solid and dashed lines represent the theoretical phase boundaries for $\varphi=0.01$ and $0.10$, respectively.
	It should be noted that the theoretical phase boundary does not exist for $\varphi=0.10$ and $\vartheta<1$.}
	\label{fig:probability}
\end{figure}

We have also evaluated from our simulations the ``probabilities'' for the NME and AME to take place.
As written before, ensemble averages over the histories of $N=100$ different initial configurations are taken for each initial condition.
Then, for $\vartheta>1$, we introduce the probability for NME as
\begin{equation}
	P_{\rm NME}(\tau)
	= \frac{1}{N^2}
	\sum_{i=1}^N \sum_{j=1}^N \Theta\left(\theta_{{\rm FS},i}(\tau) - \theta_{{\rm FQE},j}(\tau)\right),	
	\label{eq:P_NME}
\end{equation}
where $\theta_{{\rm FS},i}(\tau)$ and $\theta_{{\rm FQE},j}(\tau)$ stand for the values of $\theta_{\rm FS}(\tau)$ and $\theta_{\rm FQE}(\tau)$ in the evolutions from the initial configurations $i$ and $j$, respectively.
Thus, given a pair of initial conditions characterized by $\dot{\gamma}^*_{\rm ini}>\dot{\gamma}^*_{\rm tar}$ and $\vartheta>1$, $P_{\rm NME}(\tau)$ measures the fraction of pair microscopic histories where at time $\tau$ the temperature of the FS system has become higher than that of the FQE system.
Similarly, for $\vartheta<1$, we introduce the probability for AME as
\begin{equation}
	P_{\rm AME}(\tau)
	= \frac{1}{N^2}
	\sum_{i=1}^N \sum_{j=1}^N
	\Theta\left(\theta_{{\rm FQE},i}(\tau) - \theta_{{\rm FS},j}(\tau)\right).\label{eq:P_AME}
\end{equation}
Figure \ref{fig:probability} represents the probabilities $P_{\rm NME}(\tau)$ (symbols in the region $\vartheta>1$) and $P_{\rm AME}(\tau)$  (symbols in the region $\vartheta<1$) at several values of $\tau$ and for $e=0.9$, $T_{\rm env}^{(\rm tar)*}=1.0$, $\dot{\gamma}_{\rm tar}^*=1.0$, $\dot{\gamma}_{\rm ini}^*=2.0$, and two values of the volume fraction ($\varphi=0.01$ and $\varphi=0.10$).
In these figures, the vertical solid and dashed lines express the phase boundaries obtained from the kinetic theory for $\varphi=0.01$ and $0.10$, respectively.
The absence of the vertical dashed line for $\vartheta<1$ means that the theory does not predict the existence of AME for  $\varphi=0.10$, as observed from Fig.\ \ref{fig:phase_xi}(b).
It is remarkable that the probability $P_{\rm NME}$ has a sharp change around the theoretical boundaries at $\vartheta=1$ for $\tau\lesssim1$, but it is almost independent of $\vartheta$ for $\tau\gtrsim 2$.
We also note that the probability $P_{\rm AME}$ is quite low in its domain for $\tau\lesssim1$, while it becomes relatively high for $\tau\gtrsim2$.
We notice that the crossing time of the temperatures vanishes on the boundary between the ``NME'' and ``No Mpemba'' regions. 
Similarly, the second crossing time is infinity on the boundary between the ``NME'' and ``NME+AME'' regions (see also Fig.\ \ref{fig:Mpemba}(a)). 

Let us indicate another characteristic point of the ``AME'' for $\vartheta<1$.
As shown in Fig.~\ref{fig:phase_xi}(a), the ``AME''  is absent  in the regime of high $\dot{\gamma}_{\rm ini}/\dot\gamma_{\rm tar}$, regardless of the value of $\vartheta<1$. 
This feature is originated by  collision effects between grains, because the AME exists even for high $\dot{\gamma}^*_{\rm ini}$ when the collisionless model is considered (see Appendix \ref{sec:collisionless}).
It should be noted that the crossing time is located at infinity on the boundary between the ``AME'' and ``No Mpemba'' regions.
In addition, the first crossing time vanishes on the boundary between the ``NME+AME'' ($\vartheta>1$) and ``AME'' ($\vartheta<1$) regions (see also Fig.\ \ref{fig:Mpemba}(b)). 

\begin{figure}[htbp]
	\includegraphics[width=0.5\linewidth]{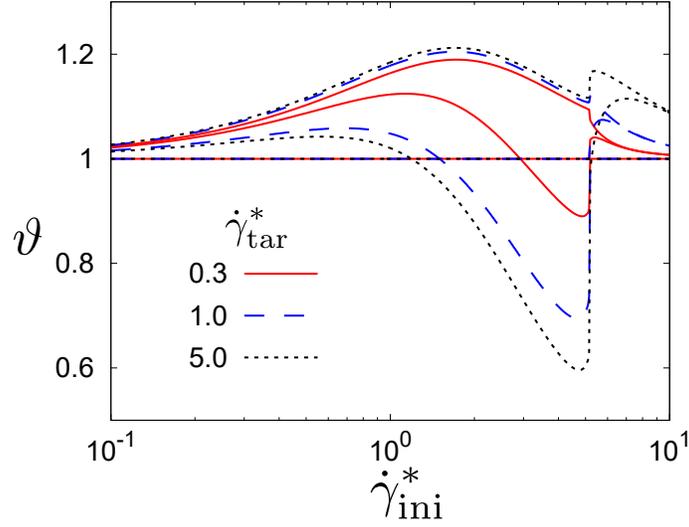}
	\caption{Phase diagram on the plane $\vartheta$ versus $\dot\gamma_{\rm ini}^*$ for $e=0.9$, $\varphi=0.01$, $T_{\rm env}^{({\rm tar})*}=1.0$, and several values of the target shear rate $\dot{\gamma}_{\rm tar}^*$.}
	\label{fig:target_shear}
\end{figure}

\begin{figure}[htbp]
	\includegraphics[width=0.5\linewidth]{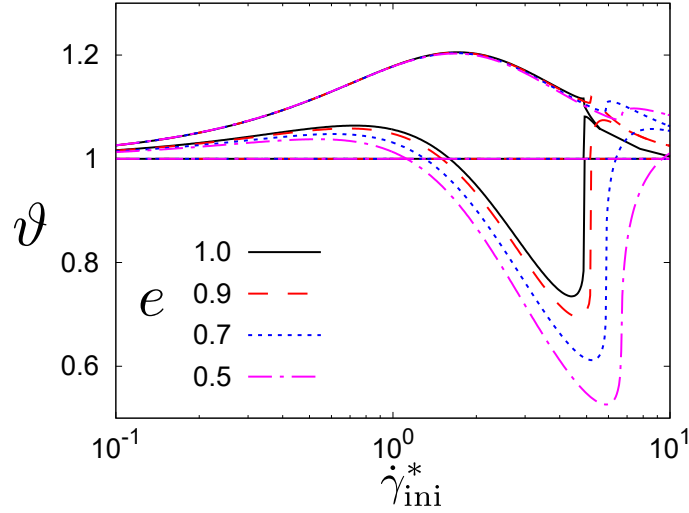}
	\caption{Phase diagram  on the plane $\vartheta$ versus $\dot\gamma_{\rm ini}^*$  for $\varphi=0.01$, $T_{\rm env}^{({\rm tar})*}=1.0$, $\dot{\gamma}^*_{\rm tar}=1.0$, and several values of the  restitution coefficient $e$.}
	\label{fig:phase_e}
\end{figure}

\begin{figure}[htbp]
	\includegraphics[width=0.5\linewidth]{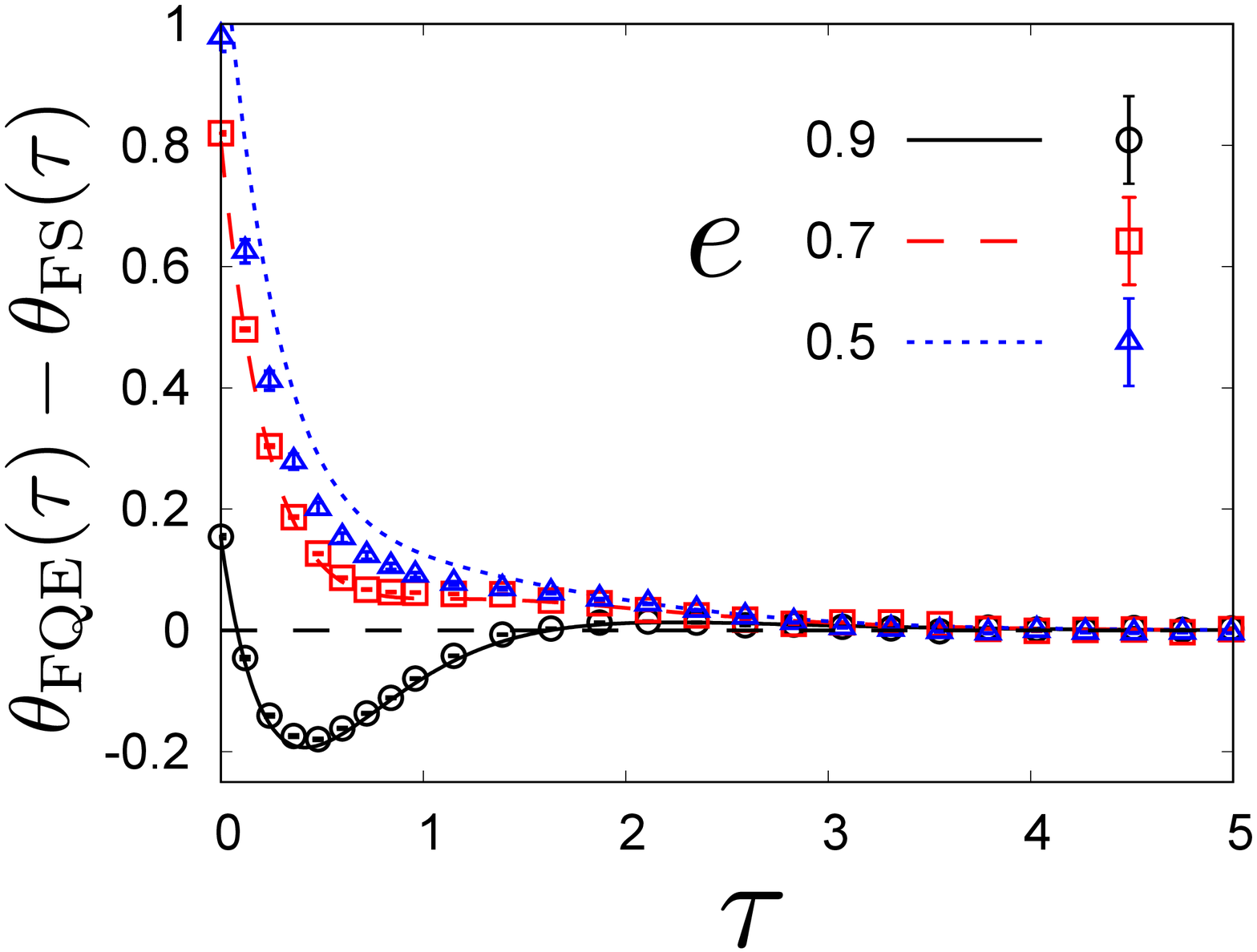}
	\caption{Time evolutions of the temperature differences $\theta_{\rm FQE}(\tau)-\theta_{\rm FS}(\tau)$ for $\varphi=0.01$, $T_{\rm env}^{({\rm ini})*}=5.29$, $T_{\rm env}^{({\rm tar})*}=1.0$, $\dot{\gamma}^*_{\rm ini}=4.0$, $\dot{\gamma}^*_{\rm tar}=1.0$, and restitution coefficients $e=0.9$, $0.7$, and $0.5$.
	The symbols are obtained from our simulations and the lines are kinetic-theory predictions.}
	\label{fig:e_comp}
\end{figure}

Thus far, we have fixed the target shear rate at $\dot{\gamma}^*_{\rm tar}=1$ and the restitution coefficient at $e=0.9$. 
Let us now explore the influence of varying $\dot{\gamma}^*_{\rm tar}$ and $e$ on the phase diagram.   Figure~\ref{fig:target_shear} shows the phase diagrams for $e=0.9$, $\varphi=0.01$, $T_{\rm env}^{(\rm tar)*}=1.0$, and three values of the target shear rate: $\dot{\gamma}^*_{\rm tar}=0.30$, $1.0$, and $5.0$.
Although the boundary between the ``NME+AME'' and ``No Mpemba'' regions seems to be rather insensitive to the choice of the target shear rate, the ``AME''  region  ($\vartheta<1$) becomes smaller as the target shear rate decreases.

The influence of the restitution coefficient is illustrated in Fig.\ \ref{fig:phase_e}. We observe that the shape of the phase boundary between the ``NME+AME'' and ``No Mpemba'' regions is practically insensitive to the restitution coefficient, while the ``AME'' region ($\vartheta<1$) shrinks as collisions become less inelastic. In any case, it is important to remark that the Mpemba effect is still clearly present in the case of elastic collisions ($e=1$), as illustrated in Fig.\ \ref{fig:Mpemba}.

As shown by Fig.\ \ref{fig:e_comp}, we note that the theory can reproduce the simulation results even in strongly inelastic cases, except for the initial relaxation for $e=0.5$.

\section{Inverse and mixed Mpemba effects}\label{Sec:Inverse_Mixed}

In this section, we develop the analysis of the inverse Mpemba effect (IME) as a counterintuitive heating effect similar to the cooling Mpemba effect.
We also discuss the mixed Mpemba effect (MME) in which both heating and cooling processes can be observed during the time evolution.
The heating process in the FS system implies the choice $\dot{\gamma}_{\rm ini}^*<\dot{\gamma}_{\rm tar}^*$.
As in Sec.\ \ref{sec:simulation},  the target environmental temperature is fixed as $T_{\rm env}^{(\rm tar)*}=1.0$.
Since the physics in IME and MME is common to that of the Mpemba effect in cooling processes, we only illustrate some examples of IME and MME in this section.

\subsection{Inverse Mpemba effect}

\begin{figure}[htbp]
	\includegraphics[width=0.8\linewidth]{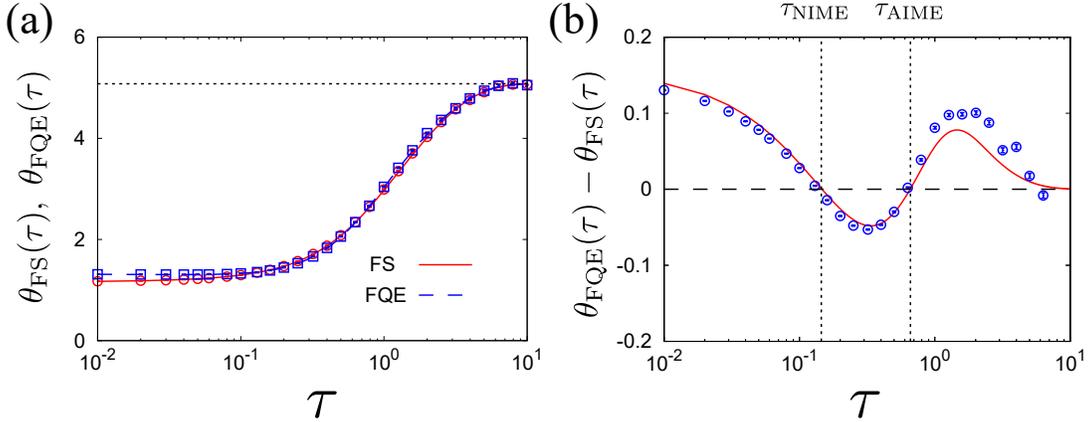}
	\caption{
Time evolutions of (a) the temperatures $\theta_{\rm FS}(\tau)$ and $\theta_{\rm FQE}(\tau)$ and (b) the temperature difference $\theta_{\rm FQE}(\tau)-\theta_{\rm FS}(\tau)$ for $e=0.9$, $\varphi=0.01$, $T_{\rm env}^{(\rm tar)*}=1.0$, $\dot{\gamma}_{\rm tar}^*=4.0$,  $\dot{\gamma}_{\rm ini}^*=1.0$, and $T_{\rm env}^{(\rm ini)*}=1.33$ as an example of AIME.
	The dotted line in panel (a) represents the target temperature $\theta_{\rm tar}=5.08$.
	The vertical dotted lines in panel (b) indicate $\tau_{\rm NIME}$ and $\tau_{\rm AIME}$.
	The symbols (with error bars) are obtained from our simulations and the lines are kinetic-theory predictions.
	Panel (b) presents a typical example of ``NIME+AIME''.}
	\label{fig:evol_xi_inverse}
\end{figure}

\begin{figure}[htbp]
	\includegraphics[width=0.5\linewidth]{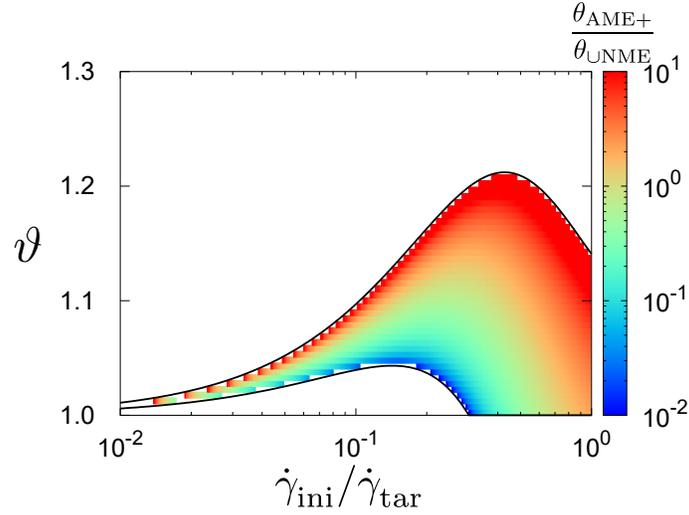}
	\caption{Magnitude plot of the ratio of the amplitudes defined in Eqs.\  \eqref{eq:theta_NME_prime&eq:theta_AME_prime} for the ``NIME+AIME'' region corresponding to $\varphi=0.01$,  $e=0.9$,  $T_{\rm env}^{(\rm tar)*}=1.0$, and $\dot{\gamma}^*_{\rm tar}=4.0$.}
	\label{fig:phase_xi_0.01_inverse}
\end{figure}

IME is a heating process in which a liquid starting from a lower initial temperature can have a higher temperature than that starting from a higher initial temperature, both initial temperatures being lower than the final steady temperature.
While this definition is simple, there might be some additional overshoots of the temperature.

A typical heating process with $\theta_{\rm FQE}(0)>\theta_{\rm FS}(0)$ for a dilute suspension ($\varphi=0.01$) with the parameters $\dot{\gamma}_{\rm ini}^*=1.0$, $\dot{\gamma}_{\rm tar}^*=4.0$, $e=0.9$, $T_{\rm env}^{(\rm tar)*}=1.0$, and $T_{\rm env}^{(\rm ini)*}=1.33$ ($\vartheta=1.13$) is shown in Fig.~\ref{fig:evol_xi_inverse}. 
It can be observed that, although $\theta_{\rm FQE}(\tau)-\theta_{\rm FS}(\tau)>0$ in the early stage $\tau<\tau_{\rm NIME}$ of the relaxation process, one has $\theta_{\rm FQE}(\tau)-\theta_{\rm FS}(\tau)<0$ for intermediate times $\tau_{\rm NIME}<\tau<\tau_{\rm AIME}$ (which corresponds to NIME), and then $\theta_{\rm FQE}(\tau)-\theta_{\rm FS}(\tau)>0$ for later times $\tau>\tau_{\rm AIME}$ (which corresponds to AIME), until finally both $\theta_{\rm FQE}(\tau)$ and $\theta_{\rm FS}(\tau)$ reach a common stationary value. 
Here, $\tau_{\rm NIME}$ and $\tau_{\rm AIME}$ are defined analogously to $\tau_{\rm NME}$ and $\tau_{\rm AME}$ in Eqs.\ \eqref{eq:tau_NME_AME_def1} and \eqref{eq:tau_NME_AME_def2}, respectively.
Thus, this is an example of a ``NIME+AIME'' process analogous to the ``NME+AME'' in cooling processes discussed in Sec.\ \ref{sec:simulation}. 
The two  amplitudes characterizing the NIME+AIME phenomenon can be defined as in Eqs.\ \eqref{eq:theta_NME_prime&eq:theta_AME_prime}.
It is remarkable that the amplitude of AIME is larger than that of NIME in Fig.~\ref{fig:evol_xi_inverse}.
As Fig.\ \ref{fig:phase_xi_0.01_inverse} shows, this property  is observed when the value of the ratio $\vartheta$ is near
the upper boundary curve, in analogy to the NME+AME case shown in Fig.\ \ref{fig:phase_xi_0.01}.


\subsection{Mixed Mpemba effect}\label{sec:mixed}

\begin{figure}[htbp]
	\includegraphics[width=0.8\linewidth]{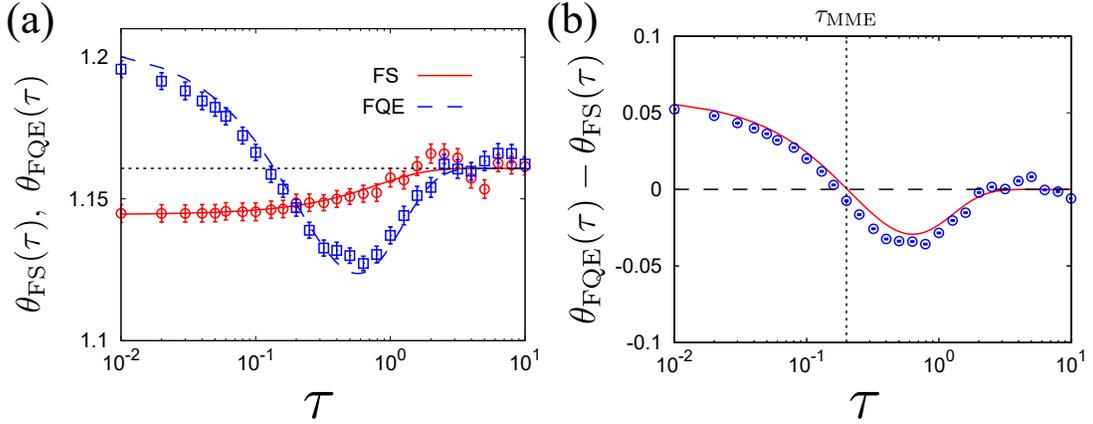}
	\caption{Time evolutions of (a) the temperatures $\theta_{\rm FS}(\tau)$ and $\theta_{\rm FQE}(\tau)$ and (b) the temperature difference $\theta_{\rm FQE}(\tau)-\theta_{\rm FS}(\tau)$ for $e=0.9$, $\varphi=0.01$, $T_{\rm env}^{(\rm tar)*}=1.0$, $\dot{\gamma}_{\rm tar}^*=1.0$,  $\dot{\gamma}_{\rm ini}^*=0.95$, and  $T_{\rm env}^{(\rm ini)*}=1.21$ for this choice of parameters,  $\vartheta=1.05$ as an example of MME.
	The dotted line in panel (a) represents the target temperature $\theta_{\rm tar}=1.16$.
	The vertical dotted line in panel (b) indicates $\tau_{\rm MME}$.
	The symbols (with error bars) are obtained from our simulations and the lines are kinetic-theory predictions.}
	\label{fig:evol_mixed}
\end{figure}

So far, we have focused on systems where both initial temperatures $\theta_{\rm FS}(0)$ and $\theta_{\rm FQE}(0)$ are either higher or lower than the final stationary temperature $\theta_{\rm tar}$ corresponding to the target quantities $\dot{\gamma}_{\rm tar}^*$ and $T_{\rm env}^{({\rm tar})*}$.
On the other hand, we can consider the case where one of the initial temperatures is higher and the  other one is lower than the target temperature $\theta_{\rm tar}$.
In conventional relaxation processes, the curves representing the temperature evolution for both systems are expected not to meet (except asymptotically in the steady state). 
However, if this is not the case, we face a situation that can be referred to as the mixed Mpemba effect (MME).
We have observed that the MME takes place only when  $\theta_{\rm FQE}(0)>\theta_{\rm tar}>\theta_{\rm FS}(0)$. 
While $\theta_{\rm FS}(\tau)$ increases monotonically, the relaxation of $\theta_{\rm FQE}(\tau)$ presents a transient behavior where it overshoots first the value $\theta_{\rm tar}$ and then the curve $\theta_{\rm FS}(\tau)$. 
This MME is illustrated in Fig.\ \ref{fig:evol_mixed} for $e=0.9$, $\varphi=0.01$, $T_{\rm env}^{(\rm tar)*}=1.0$, $\dot{\gamma}_{\rm tar}^*=1.0$,  $\dot{\gamma}_{\rm ini}^*=0.95$, and $T_{\rm env}^{(\rm ini)*}=1.21$.
Here, we define $\tau_{\rm MME}$ which satisfies both $\theta_{\rm FQE}(\tau_{\rm MME})-\theta_{\rm FS}(\tau_{\rm MME})=0$ and $\theta_{\rm FQE}(\tau)-\theta_{\rm FS}(\tau)>0$ for $\tau<\tau_{\rm MME}$.
It is interesting to note that Fig.\ \ref{fig:evol_mixed}(b) is reminiscent of Figs.\ \ref{fig:evol_xi}(b), (d) and \ref{fig:evol_xi_inverse}(b), except that there is only one crossing, so that the last stage where $\theta_{\rm FQE}(\tau)-\theta_{\rm FS}(\tau)$ becomes positive again is missing in the kinetic theory as shown in Fig.\ \ref{fig:evol_mixed}(b) as far as we have checked.
On the other hand, we observe the second and even the third crossings in the later stage in our simulations.
We should clarify the origin of this discrepancy between the simulation and kinetic theory in the near future.


\section{Discussion and conclusion}\label{sec:discussion}

In this paper, we have demonstrated by kinetic theory and computer simulations that the Mpemba effect can be observed in sheared inertial suspensions.
We have also illustrated that there are two classes of Mpemba effects, NME and AME; NME is generic (the initially hotter suspension starts cooling down  more rapidly than the initially colder suspension, eventually catching up the latter), while AME is nontrivial (even though the initially hotter suspension starts cooling down  more slowly than the initially colder suspension, the former eventually catches up the latter).
NME can be observed if we compare the transient dynamics of a system starting from a ``hot'' (unsheared) quasi-equilibrium state with that of a system starting from a ``cold''  nonequilibrium sheared steady state. This is because the initial cooling rate is higher in the former due to the absence of viscous heating than in the latter.
AME can be observed as the transient dynamics to approach a common steady state when the initial quasi-equilibrium state is colder than the initial  nonequilibrium sheared state.
Interestingly, a double crossing (NME followed by AME) can  also be observed. 

Similarly, we have confirmed the existence of NIME and AIME in the (heating) inverse  Mpemba effect, as well  a mixed Mpemba effect (MME) where both cooling and heating are present.
Thus, we have clarified the generic features of the Mpemba effect, which can be observed in cooling, heating, and mixed relaxation processes.
As far as we have studied, AME and NME+AME (including their inverse counterparts) are restricted to dilute inertial suspensions, while NME (including NIME and MME) are present for moderately dense systems.

Although the main text is dedicated to  exotic relaxation processes of temperature, a similar exotic relaxation process can be observed in the measurement of the viscosity.
This point is addressed in Appendix \ref{sec:eta}, where it turns out that the viscosity difference changes its sign at most once, as far as we have checked.

It must be emphasized that the model we have analyzed in this paper is oversimplified in what the treatment of the hydrodynamic interactions between particles.
Thus, the investigation of more realistic suspensions will be a future subject of our study.

In particular, a recent experimental study on the Mpemba effect for a single colloid particle in water trapped in a potential created by optical tweezers~\cite{Kumar20, Chetrite21} is interesting
in that (i)  a collisionless model without consideration of hydrodynamic interactions is used to explain the experiments,
and (ii) the main feature of Mpemba effect can be understood by hopping processes from one local minimum to another minimum in the free energy.
Because we have not taken into account the local potential trapping effects, it would be interesting to include such effects in the future.

\acknowledgements

We thank Vicente Garz\'{o} for useful comments.
The work of S.T.\ is partially supported by the Grant-in-Aid of MEXT for Scientific Research (Grant No.\ 20K14428).
The work of H.H.\ is partially supported by ISHIZUE 2020 of Kyoto University Research Development Program.
The research of A.S.\ has been supported by the Spanish Agencia Estatal de Investigaci\'on through Grant No.\ FIS2016-76359-P and by the Junta de Extremadura (Spain) through Grant No.\ GR18079, both partially financed by FEDER funds.
The authors thank YITP activity YITP--X--19--01, and S.T.\ and H.H.\ also thank another YITP activity YITP-T--18--03.
S.T.\ and H.H.\ appreciate the warm hospitality of the University of Extremadura during their stays.
A.S.\ appreciates the warm hospitality of the Yukawa Institute for Theoretical Physics, Kyoto University during his stay.


\appendix

\section{Moment equations}\label{sec:derivation}
Let us write in this Appendix the explicit expressions of the moment equations.
The equation of the kinetic stress $P^k_{\alpha\beta}$ can be obtained by multiplying both sides of Eq.\ \eqref{Enskog} by $m V_\alpha V_\beta$ and integrating over $\bm{V}$. The result is
\begin{equation}
	\frac{\partial}{\partial t}P^k_{\alpha\beta}
	+\dot{\gamma} (\delta_{\alpha x}P_{y \beta}^k+\delta_{\beta x} P_{y \alpha}^k)
	=-2\zeta ( P_{\alpha\beta}^k- n T_{\rm env} \delta_{\alpha\beta} )
	-\Lambda_{\alpha\beta},
	\label{Garzo31}
\end{equation}
where
\begin{equation}
	\label{Garzo32}
	\Lambda_{\alpha\beta} \equiv -m \int d\bm{V} V_\alpha V_\beta J_{\rm E}[\bm{V}|f,f].
\end{equation}
The time-dependent equations for $T$, $\Delta T$, $\delta T$, and $P_{xy}^k$ can be easily derived from Eq.\
\eqref{Garzo31}.
They are given by
\begin{subequations}
\label{partT-part_P_{xy}}
\begin{eqnarray}
	\label{partT}
	\frac{\partial}{\partial t} T&=&
	-\frac{2}{3 n}\dot{\gamma} P^k_{xy}+2\zeta (T_{\rm env}-T) - \frac{{\Lambda}_{\alpha\alpha}}{3n},\\
	\label{part_DT}
	\frac{\partial}{\partial t} \Delta T&=&
	-\frac{2}{n}\dot{\gamma} P^k_{xy}-2\zeta \Delta T-\frac{{\Lambda}_{xx}-{\Lambda}_{yy}}{n},\\
	\label{part_dT}
	\frac{\partial}{\partial t}\delta T&=&
	-\frac{2}{n}\dot{\gamma} P^k_{xy}-2\zeta \delta T-\frac{{\Lambda}_{xx}-{\Lambda}_{zz}}{n},\\
	\label{part_P_{xy}}
	\frac{\partial}{\partial t}P_{xy}^k&=&- \dot{\gamma} P^k_{yy}
	-2\zeta P_{xy}^k-{\Lambda}_{xy}.
\end{eqnarray}
\end{subequations}

Note that ${\Lambda}_{\alpha\beta}$ in Eq.~\eqref{Garzo32} can be expressed as~\cite{Hayakawa17,Takada20a, Santos98, Montanero99}
\begin{equation}\label{eq:overline_Lambda^E}
	{\Lambda}_{\alpha\beta}
	= \frac{1+e}{4} m \sigma^{2} g_0
	[L_{\alpha\beta}+(1-e) M_{\alpha\beta}],
\end{equation}
where we have introduced
\begin{subequations}
\label{Lab-Mab}
\begin{align}
\label{Lab}
	L_{\alpha \beta} &\equiv
	\int d\widehat{\bm{\sigma}}
	\left[
	\widehat{\sigma}_\alpha I_\beta^{(2)}\left(\widehat{\bm{\sigma}}\right) + \widehat{\sigma}_\beta I_\alpha^{(2)}\left(\widehat{\bm{\sigma}}\right)
		-2\widehat{\sigma}_\alpha \widehat{\sigma}_\beta I^{(3)}\left(\widehat{\bm{\sigma}}\right)
+a\left(\delta_{\alpha x} \widehat{\sigma}_\beta + \delta_{\beta x} \widehat{\sigma}_\alpha\right)I^{(2)}\left(\widehat{\bm{\sigma}}\right)
	\right],\\
	\label{Mab}
M_{\alpha \beta} &\equiv
	\int d\widehat{\bm{\sigma}} \widehat{\sigma}_\alpha \widehat{\sigma}_\beta I^{(3)}\left(\widehat{\bm{\sigma}}\right),
\end{align}
\end{subequations}
with $I_\alpha^{(\ell)}(\widehat{\bm{\sigma}})$ and $I^{(\ell)}\left(\widehat{\bm{\sigma}}\right)$ being
\begin{subequations}
\begin{align}\label{eq:I_alpha}
	I_\alpha^{(\ell)}\left(\widehat{\bm{\sigma}}\right)
	&\equiv \int d\bm{V}_1 \int d\bm{V}_2
	\Theta\left(\widehat{\bm{\sigma}}\cdot \bm{V}_{12}\right) \left(\widehat{\bm{\sigma}}\cdot \bm{V}_{12}\right)^\ell V_{12,\alpha}
	f(\bm{V}_1 ) f(\bm{V}_2- \dot{\gamma} \sigma \widehat{\sigma}_y \bm{e}_x),\\
	I^{(\ell)}\left(\widehat{\bm{\sigma}}\right)
	&\equiv \int d\bm{V}_1 \int d\bm{V}_2
	\Theta\left(\widehat{\bm{\sigma}}\cdot \bm{V}_{12}\right) \left(\widehat{\bm{\sigma}}\cdot \bm{V}_{12}\right)^\ell
	f(\bm{V}_1 ) f(\bm{V}_2- \dot{\gamma} \sigma \widehat{\sigma}_y \bm{e}_x).
\label{eq:I}
\end{align}
\end{subequations}
Analogously, $P_{\alpha\beta}^c$ introduced in Eq.~\eqref{pressure:collisional} can be expressed as
\begin{equation}
	P_{\alpha\beta}^c
	= \frac{1+e}{4}m\sigma^3 g_0 \int d\widehat{\bm{\sigma}} \widehat{\sigma}_\alpha \widehat{\sigma}_\beta I^{(2)}\left(\widehat{\bm{\sigma}}\right) .
	\label{eq:Pc}
\end{equation}

Since $\widehat{\sigma}_\alpha I_\alpha^{(\ell)}(\widehat{\bm{\sigma}})=I^{(\ell+1)}(\widehat{\bm{\sigma}})$ is satisfied, we have the following simple relations for $e=1$:
\begin{equation}\label{eq:e=1}
\Lambda_{\alpha\alpha}=2\dot{\gamma} P_{xy}^c,
\quad
\frac{\partial}{\partial t} T=
	-\frac{2\dot{\gamma}}{3 n}P_{xy}+2\zeta (T_{\rm env}-T) ,
\end{equation}
which represents the microscopic basis of Eq.~\eqref{eq_of_T} with  $c_V=3/2$.

The explicit expression of the collision integral ${\Lambda}_{\alpha\beta}$ cannot be obtained because it needs information on the distribution function.
A good estimate of this collisional moment can be obtained by using Grad's
approximation~\cite{Garzo13,BGK2016,Chamorro15,Grad49,Garzo02,Santos04}
\begin{equation}
	\label{Grad}
	f(\bm{V})=f_{\rm eq}(\bm{V})\left(1+\frac{m}{2T}\Pi_{\alpha\beta}V_\alpha V_\beta\right),
\end{equation}
where
\begin{equation}
	\label{Maxwell}
	f_{\rm eq}(\bm{V})=
	n\left(\frac{m}{2\pi T}\right)^{3/2} \exp\left(-\frac{m V^2}{2T} \right)
\end{equation}
is the Maxwellian distribution and
\begin{equation}
	\label{vic5}
	\Pi_{\alpha\beta} \equiv \frac{P^k_{\alpha\beta}}{nT}-\delta_{\alpha\beta}
\end{equation}
is the traceless part of the (dimensionless) kinetic pressure tensor $P^k_{\alpha\beta}$.

In Grad's approximation, the dimensionless collisional moments $\Lambda_{\alpha\beta}^*\equiv \Lambda_{\alpha\beta}/(n\zeta T_{\rm env})$ and collisional shear stress $\Pi_{xy}^{c*}\equiv P_{xy}^c/(nT_{\rm env})$ are given by~\cite{Takada20a,Santos98, Montanero99}
\begin{subequations}
\label{Lambda*}
\begin{align}
	\Lambda_{\alpha\beta}^*
	&=\frac{3\sqrt{2}}{\pi}(1+e)\varphi g_0 \sqrt{T_{\rm env}^*}\theta^{3/2}
		\int d\widehat{\bm{\sigma}}
	\left[\widehat{\sigma}_\alpha \hat{J}_\beta(\widehat{\bm{\sigma}}) + \widehat{\sigma}_\beta \hat{J}_\alpha(\widehat{\bm{\sigma}})
		+(1-e)\widehat{\sigma}_\alpha \widehat{\sigma}_\beta \hat{I}^{(3)}(\widehat{\bm{\sigma}})
		+2b_T(\widehat{\bm{\sigma}}) \widehat{\sigma}_\alpha \widehat{\sigma}_\beta \hat{I}^{(2)}(\widehat{\bm{\sigma}})\right],\label{eq:Lambda*_alpha_beta}\\
	\Lambda_{\alpha\alpha}^*
	&=\frac{3\sqrt{2}}{\pi}(1+e)\varphi g_0 \sqrt{T_{\rm env}^*} \theta^{3/2}
		\int d\widehat{\bm{\sigma}}
		\left[(1-e)\hat{I}^{(3)}(\widehat{\bm{\sigma}}) +2b_T(\widehat{\bm{\sigma}}) \hat{I}^{(2)}(\widehat{\bm{\sigma}})\right],\\
    \Pi_{xy}^{c*}&=\frac{3}{\pi}(1+e)\varphi g_0 \theta
		\int d\widehat{\bm{\sigma}}
	\widehat{\sigma}_x \widehat{\sigma}_y \hat{I}^{(2)}(\widehat{\bm{\sigma}}),\label{eq:Pxyc*}
\end{align}
\end{subequations}
where we have introduced the function
\begin{equation}
\label{bT}
	b_T(\widehat{\bm{\sigma}})\equiv \frac{\tilde{\dot{\gamma}}}{\sqrt{2}} \widehat{\sigma}_x \widehat{\sigma}_y,\quad \tilde{\dot{\gamma}}\equiv \frac{\dot{\gamma}\sigma}{\sqrt{T/m}}=\frac{\dot{\gamma}^*}{\sqrt{T_{\rm env}^* \theta}}.
\end{equation}
The quantities $\hat{I}^{(2)}$, $\hat{I}^{(3)}$, and $\hat{J}_\alpha$ are given by \cite{Takada20a,Santos98, Montanero99}
\begin{subequations}
\label{I2,I3,J}
\begin{align}
	\hat{I}^{(2)}(\widehat{\bm{\sigma}})
	&= -\frac{b_T}{\sqrt{2\pi}}e^{-b_T^2/2} + \frac{1+b_T^2}{2} {\rm erfc}\left(\frac{b_T}{\sqrt{2}}\right)
	+\frac{1}{2} {\rm erfc}\left(\frac{b_T}{\sqrt{2}}\right) \widehat{\sigma}_\beta \widehat{\sigma}_\gamma \Pi_{\beta \gamma}
	+\frac{b_T}{8\sqrt{2\pi}}e^{-b_T^2/2}\left(\widehat{\sigma}_\beta \widehat{\sigma}_\gamma \Pi_{\beta \gamma}\right)^2,\\
	\hat{I}^{(3)}(\widehat{\bm{\sigma}})
	&= \frac{2+b_T^2}{\sqrt{2\pi}}e^{-b_T^2/2} -\frac{1}{2}b_T \left(3+b_T^2\right) {\rm erfc}\left(\frac{b_T}{\sqrt{2}}\right)
	+3\left[\frac{e^{-b_T^2/2}}{\sqrt{2\pi}} - \frac{b_T}{2}{\rm erfc}\left(\frac{b_T}{\sqrt{2}}\right)\right]
	\widehat{\sigma}_\beta \widehat{\sigma}_\gamma \Pi_{\beta \gamma}\nonumber\\
	&\hspace{1em}+\frac{3}{8\sqrt{2\pi}}e^{-b_T^2/2}\left(\widehat{\sigma}_\beta \widehat{\sigma}_\gamma \Pi_{\beta \gamma}\right)^2,\\
	\hat{J}_x (\widehat{\bm{\sigma}})
	&= -\widehat{\sigma}_x
	\left(\widehat{\sigma}_\beta \widehat{\sigma}_\gamma \Pi_{\beta\gamma} - \Pi_{xx} - \frac{\widehat{\sigma}_y}{\widehat{\sigma}_x}\Pi_{xy}\right)
	\left[\sqrt{\frac{2}{\pi}} e^{-b_T^2/2} - b_T {\rm erfc}\left(\frac{b_T}{\sqrt{2}}\right)
	+\frac{1}{2\sqrt{2\pi}}e^{-b_T^2/2} \widehat{\sigma}_\beta \widehat{\sigma}_\gamma \Pi_{\beta\gamma}\right],\\
	\hat{J}_y (\widehat{\bm{\sigma}})
	&= -\widehat{\sigma}_y
	\left(\widehat{\sigma}_\beta \widehat{\sigma}_\gamma \Pi_{\beta\gamma} - \Pi_{yy} - \frac{\widehat{\sigma}_x}{\widehat{\sigma}_y}\Pi_{xy}\right)
	\left[\sqrt{\frac{2}{\pi}} e^{-b_T^2/2} - b_T {\rm erfc}\left(\frac{b_T}{\sqrt{2}}\right)
	+\frac{1}{2\sqrt{2\pi}}e^{-b_T^2/2} \widehat{\sigma}_\beta \widehat{\sigma}_\gamma \Pi_{\beta\gamma}\right],\\
    \hat{J}_z (\widehat{\bm{\sigma}})
	&= -\widehat{\sigma}_z
	\left(\widehat{\sigma}_\beta \widehat{\sigma}_\gamma \Pi_{\beta\gamma} - \Pi_{zz} \right)
	\left[\sqrt{\frac{2}{\pi}} e^{-b_T^2/2} - b_T {\rm erfc}\left(\frac{b_T}{\sqrt{2}}\right)
	+\frac{1}{2\sqrt{2\pi}}e^{-b_T^2/2} \widehat{\sigma}_\beta \widehat{\sigma}_\gamma \Pi_{\beta\gamma}\right],
\end{align}
\end{subequations}
where we have used the complementary error function ${\rm erfc}(x)\equiv (2/\sqrt{\pi})\int_x^\infty dt e^{-t^2}$.

Thus, the dimensionless time evolution equations of the temperature, the two temperature differences, and the kinetic shear stress are, respectively, given by
\begin{subequations}
\label{eq:evol_theta-eq:evol_Pi_xy}
\begin{align}
	\partial_\tau \theta
	&=  - \frac{2}{3}\dot{\gamma}^* \Pi_{xy}^*-2(\theta-1)-\frac{1}{3}\Lambda_{\alpha\alpha}^*,\label{eq:evol_theta}\\
	\partial_\tau \Delta \theta
	&= -2\dot{\gamma}^* \Pi_{xy}^* -2\Delta\theta -\delta \Lambda_{xx}^*+\delta \Lambda_{yy}^*,\\
	\partial_\tau \delta \theta
	&= -2\dot{\gamma}^* \Pi_{xy}^* -2\delta\theta-2\delta \Lambda_{xx}^*-\delta\Lambda_{yy}^*,\\
	\partial_\tau \Pi_{xy}^*
	&=  - \dot{\gamma}^* \left(\theta-\frac{2}{3}\Delta\theta + \frac{1}{3}\delta\theta\right)-2\Pi_{xy}^*-\Lambda_{xy}^*,\label{eq:evol_Pi_xy}
\end{align}
\end{subequations}
with $\Pi^*_{\alpha\beta}\equiv P^k_{\alpha\beta}/(nT_{\rm env})-\theta\delta_{\alpha\beta}=\theta\Pi_{\alpha\beta}$, $\Delta\theta\equiv (P_{xx}^k - P_{yy}^k)/(nT_{\rm env})$, $\delta\theta\equiv (P_{xx}^k-P_{zz}^k)/(nT_{\rm env})$, and 
\begin{equation}
	\delta \Lambda_{xx}^*=\Lambda_{xx}^*-\frac{1}{3}\Lambda_{\alpha\alpha}^*,\quad
	\delta \Lambda_{yy}^*=\Lambda_{yy}^*-\frac{1}{3}\Lambda_{\alpha\alpha}^*.
\end{equation}
Equations \eqref{eq:evol_theta-eq:evol_Pi_xy}, complemented by Eqs.\ \eqref{Lambda*} and \eqref{I2,I3,J}, make a closed set of coupled differential equations that can be numerically solved starting from any given initial condition.
However, due to the fact that the integrations over $\widehat{\bm{\sigma}}$ in Eqs.\ \eqref{Lambda*} need to be performed numerically at each time $\tau$, the time-dependent numerical solution of the set of evolution equations \eqref{eq:evol_theta-eq:evol_Pi_xy} is rather time-consuming. Following Ref.~\cite{Takada20a},
 this technical problem is easily overcome if, instead of the full nonlinear dependence of $\Lambda_{\alpha\beta}^*$ on the shear rate, one expands those quantities in powers of the dimensionless parameter $\tilde{\dot{\gamma}}$ defined in Eq.\ \eqref{bT}.
The quantities $\Lambda_{\alpha\alpha}^*$, $\Lambda_{xy}^*$, $\delta\Lambda_{xx}^*$, $\delta\Lambda_{yy}^*$, and $\Pi_{xy}^{c*}$ are, respectively, expanded as
\begin{subequations}
\begin{align}
	\Lambda_{\alpha\alpha}^*
	&= \varphi g_0 \sqrt{T_{\rm env}^*}\theta^{3/2}\sum_{i=0}^{N_c}
		\widetilde{\Lambda}_{\alpha\alpha}^{(i)*}\tilde{\dot{\gamma}}^{i},\label{eq:Lii_exp}\\
	\Lambda_{xy}^*
	&= \varphi g_0 \sqrt{T_{\rm env}^*}\theta^{3/2}\sum_{i=0}^{N_c}
		\widetilde{\Lambda}_{xy}^{(i)*}\tilde{\dot{\gamma}}^{i},\\
	\delta\Lambda_{xx}^*
	&= \varphi g_0 \sqrt{T_{\rm env}^*}\theta^{3/2}\sum_{i=0}^{N_c}
		\delta\widetilde{\Lambda}_{xx}^{(i)*}\tilde{\dot{\gamma}}^{i},\\
	\delta\Lambda_{yy}^*
	&= \varphi g_0 \sqrt{T_{\rm env}^*}\theta^{3/2}\sum_{i=0}^{N_c}
		\delta\widetilde{\Lambda}_{yy}^{(i)*}\tilde{\dot{\gamma}}^{i},\label{eq:Lyy_exp}\\
\Pi_{xy}^{c*}
	&=\varphi g_0 \theta \sum_{i=0}^{N_c} \widetilde{\Pi}_{xy}^{c(i)*}\tilde{\dot{\gamma}}^{i}, \label{eq:Pc_exp}
\end{align}
\end{subequations}
where a truncation at order $N_c$ has been introduced.
The coefficients of the expansions are functions of $\theta$ and $\Pi_{\alpha\beta}^*$ that can be analytically evaluated term by term. The coefficients $\widetilde{\Lambda}_{\alpha\alpha}^{(i)*}$, $\widetilde{\Lambda}_{xy}^{(i)*}$, $\delta\widetilde{\Lambda}_{xx}^{(i)*}$, and $\delta\widetilde{\Lambda}_{yy}^{(i)*}$ up to $i=6$ are listed in Table I of Ref.~\cite{Takada20a}, while the coefficients $\widetilde{\Pi}_{xy}^{c(i)*}$ up to $i=6$ are listed in Table II of Ref.~\cite{Takada20a}.
As has been demonstrated in Ref.~\cite{Takada20a}, a truncation order $N_c=6$ gives convergent results very close to those obtained from the full nonlinear expressions (formally equivalent to $N_c\to\infty$).
Thus, henceforth we adopt the sixth-order expansions, i.e. $N_c=6$, and numerically solve the time evolutions of the quantities $\theta$, $\Delta\theta$, $\delta\theta$, and $\Pi_{xy}^*$, as shown in Secs.\ \ref{sec:simulation} and \ref{Sec:Inverse_Mixed}.

\section{Exact results of a collisionless model}\label{sec:collisionless}

In this Appendix, we discuss the Mpemba effect for a collisionless model.
Because the hydrodynamic lubrication force between particles prevents particles from collisions, this model might be more realistic than the collisional model discussed in the main text.
We also compare the theoretical results of the collisionless model with those for very dilute collisional systems.

In the collisionless model, the time evolutions of the stress tensor and the kinetic temperature are obtained from Eqs.\ \eqref{eq:evol_theta-eq:evol_Pi_xy} by setting the collisional moments $\Lambda^*_{\alpha\beta}\to 0$. This formally corresponds to the limit $\nu\ll \zeta$ (where $\nu \propto g_0\varphi\sqrt{T/m}/\sigma$ is the collision frequency), i.e., the limit $g_0\varphi\sqrt{T^*_{\rm env}\theta}\to 0$. Thus, the moment equations become
\begin{subequations}
\label{eq:set_1-4}
\begin{align}
	\partial_\tau \theta
	&=  - \frac{2}{3}\dot{\gamma}^* \Pi_{xy}^*-2(\theta-1),\\
	\partial_\tau \Delta \theta
	&= -2\dot{\gamma}^* \Pi_{xy}^* -2\Delta\theta ,\label{eq:set_2}\\
	\partial_\tau \delta \theta
	&= -2\dot{\gamma}^* \Pi_{xy}^* -2\delta\theta,\label{eq:set_3}\\
	\partial_\tau \Pi_{xy}^*
	&=  - \dot{\gamma}^* \left(\theta-\frac{2}{3}\Delta\theta + \frac{1}{3}\delta\theta\right)-2\Pi_{xy}^*.
\end{align}
\end{subequations}
From Eqs.\ \eqref{eq:set_2} and \eqref{eq:set_3}, we obtain $\Delta\theta=\delta\theta$ if their initial values are identical.
Under this assumption, which is always valid for the FQE system and for the collisionless FS system, we can rewrite the evolution equations in matrix form as
\begin{equation}
	{\partial_\tau}
	\begin{pmatrix}
		\theta \\ \Delta\theta \\ \Pi_{xy}^*
	\end{pmatrix}
	=
	\begin{pmatrix}
		-2 & 0 & -\frac{2}{3}\dot{\gamma}^*\\
		0 & -2 & -2\dot{\gamma}^* \\
		-\dot{\gamma}^* & \frac{1}{3}\dot{\gamma}^* & -2
	\end{pmatrix}
	\begin{pmatrix}
		\theta \\ \Delta\theta \\ \Pi_{xy}^*
	\end{pmatrix}
	+
	\begin{pmatrix}
		2 \\ 0 \\ 0
	\end{pmatrix}.
	\label{eq:evol_collisionless}
\end{equation}
Introducing the steady-state solution
\begin{align}
\label{eq:steady_collisionless}
	&\theta_{\rm s} \equiv 1+\frac{1}{6}\dot{\gamma}^{*2},\quad
	\Delta\theta_{\rm s} =\delta\theta_{\rm s}\equiv \frac{1}{2}\dot{\gamma}^{*2},\quad
	\Pi_{xy,{\rm s}}^* \equiv -\frac{1}{2}\dot{\gamma}^*,
\end{align}
Eq.\ \eqref{eq:evol_collisionless} is rewritten as
\begin{equation}
	\partial_\tau\bm{x}=\mathsf{A}\cdot\bm{x},\label{eq:time_evol_x}
\end{equation}
with
\begin{equation}\label{eq:x_A}
	\bm{x}\equiv
	\begin{pmatrix}
		\theta-\theta_{\rm s} \\ \Delta\theta - \Delta\theta_{\rm s} \\ \Pi_{xy}^*-\Pi_{xy,{\rm s}}^*
	\end{pmatrix},\quad
	\mathsf{A} \equiv
	\begin{pmatrix}
		-2 & 0 & -\frac{2}{3}\dot{\gamma}^*\\
		0 & -2 & -2\dot{\gamma}^* \\
		-\dot{\gamma}^* & \frac{1}{3}\dot{\gamma}^* & -2
	\end{pmatrix}.
\end{equation}
The solution of Eq.\ \eqref{eq:time_evol_x} is
\begin{equation}
\label{sol_x}
	\bm{x}(\tau)= \exp(\tau \mathsf{A}) \cdot\bm{x}(0).
\end{equation}

The matrix $\mathsf{A}$ can be transformed to the Jordanian form in terms of
\begin{equation}\label{eq:UU^{-1}}
	\mathsf{U}=
	\begin{pmatrix}
		\frac{1}{3} & 0 & \frac{1}{2\dot{\gamma}^{*2}} \\
		1 & 0 & 0 \\
		0 & -\frac{1}{2\dot{\gamma}^*} & 0
	\end{pmatrix},\quad
	\mathsf{U}^{-1}=
	\begin{pmatrix}
		0 & 1 & 0 \\
		0 & 0 & -2\dot{\gamma}^* \\
		2\dot{\gamma}^{*2} & -\frac{2}{3}\dot{\gamma}^{*2} & 0
	\end{pmatrix}.
\end{equation}
The Jordan normal form $\mathsf{J}$ of the matrix $\mathsf{A}$ becomes
\begin{equation}
	\mathsf{J} = \mathsf{U}^{-1}\cdot\mathsf{A}\cdot\mathsf{U}=
	\begin{pmatrix}
		-2 & 1 & 0\\
		0 & -2 & 1\\
		0 & 0 & -2
	\end{pmatrix}
\end{equation}
and, therefore,
\begin{equation}
	\exp(\tau \mathsf{A})= \mathsf{U}\cdot\exp(\tau \mathsf{J})\cdot \mathsf{U}^{-1}.
\end{equation}
It can be proved by recursion that
\begin{equation}
\mathsf{J}^k=\begin{pmatrix}
		(-2)^k & k(-2)^{k-1} & -k(k-1)(-2)^{k-3}\\
		0 & (-2)^k & k(-2)^{k-1}\\
		0 & 0 & (-2)^k
	\end{pmatrix},
\end{equation}
so that
\begin{align}
	\exp(\tau \mathsf{J}) =&	
	\begin{pmatrix}
		1 & \tau & \frac{\tau^2}{2}\\
		0 & 1 & \tau\\
		0 & 0 & 1
	\end{pmatrix}
	e^{-2\tau}.
\end{align}
Consequently,
\begin{align}
	\exp(\tau \mathsf{A}) =&	
	\begin{pmatrix}
		1+\frac{1}{3}\dot{\gamma}^{*2}\tau^2 & -\frac{1}{9}\dot{\gamma}^{*2}\tau^2 & -\frac{2}{3}\dot{\gamma}^{*}\tau\\
		\dot{\gamma}^{*2}\tau^2 & 1-\frac{1}{3}\dot{\gamma}^{*2}\tau^2 & -2\dot{\gamma}^{*}\tau\\
		-\dot{\gamma}^{*}\tau & \frac{1}{3}\dot{\gamma}^{*}\tau & 1
	\end{pmatrix}
	e^{-2\tau}.
\end{align}
Thus, Eq.\ \eqref{sol_x} finally yields
\begin{subequations}
\begin{align}
	\label{theta_appA}
	\theta(\tau)
	&= \theta_{\rm s}
		+\left\{\theta(0)-\theta_{\rm s}
		-\frac{2}{3}\left[\Pi_{xy}^*(0)-\Pi_{xy,{\rm s}}^*\right]\dot{\gamma}^* \tau
		+\frac{1}{3}\left[\theta(0)-\frac{1}{3}\Delta\theta(0)-1\right]\dot{\gamma}^{*2}\tau^2\right\}e^{-2\tau},\\
	\label{Dtheta_appA}
	\Delta\theta(\tau)
	&= \Delta\theta_{\rm s} +\left\{\Delta\theta(0)-\Delta\theta_{\rm s}
		-2\left[\Pi_{xy}^*(0)-\Pi_{xy,{\rm s}}^*\right]\dot{\gamma}^*\tau
		+\left[\theta(0)-\frac{1}{3}\Delta\theta(0)-1\right]\dot{\gamma}^{*2}\tau^2\right\}e^{-2\tau},\\
	\label{Pi_appA}
	\Pi_{xy}^*(\tau)
	&= \Pi_{xy,{\rm s}}^* + \left\{\Pi_{xy}^*(0)-\Pi_{xy,{\rm s}}^*
		-\left[\theta(0)-\frac{1}{3}\Delta\theta(0)-1\right]\dot{\gamma}^*\tau\right\}e^{-2\tau},
\end{align}
\end{subequations}
where we have taken into account $\theta_{\rm s}-\frac{1}{3}\Delta\theta_{\rm s}=1$.

It should be noted that $\theta(\tau)$ and $\Delta \theta(\tau)$ in Eqs.~\eqref{theta_appA} and \eqref{Dtheta_appA} contain terms quadratic in $\tau$, while $\Pi_{xy}^*(\tau)$ in Eq.~\eqref{Pi_appA} contains a linear function of $\tau$.
The quadratic function of $\tau$ in Eq.~\eqref{theta_appA} may lead to two changes of sign in temperature differences (giving rise to  NME+AME), as will be shown below. On the other hand, the stress or the viscosity differences can have only one  change of sign.

Let us now explore the phase boundaries of the Mpemba effect. As in the main text,  we particularize to the FS and FQE systems (see Fig.\ \ref{fig:protocolB}). According to Eq.\ \eqref{eq:steady_collisionless},
the initial conditions for both systems are given by
\begin{equation}
\begin{cases}
	\displaystyle \theta_{\rm FS}(0)=\theta_0 \equiv 1+\frac{1}{6}\dot{\gamma}_{\rm ini}^{*2},\quad
	\Delta\theta_{\rm FS}(0)=\delta\theta_{\rm FS}(0)=\frac{1}{2}\dot{\gamma}_{\rm ini}^{*2},\quad
	\Pi_{xy}^{{\rm FS}*}(0)=-\frac{1}{2}\dot{\gamma}_{\rm ini}^*,\\
	\displaystyle \theta_{\rm FQE}(0)=\vartheta \theta_0,\quad
	\Delta\theta_{\rm FQE}(0)=\delta\theta_{\rm FQE}(0)=0,\quad
	\Pi_{xy}^{{\rm FQE}*}(0)=0.
\end{cases}
\end{equation}
From Eq.\ \eqref{theta_appA}, we then  obtain
\begin{align}
	&\theta_{\rm FQE}(\tau)-\theta_{\rm FS}(\tau)
	=\left[(\vartheta-1)\theta_0
	-\frac{1}{3}\dot{\gamma}_{\rm ini}^* \dot{\gamma}_{\rm tar}^*\tau
	+\frac{1}{3}\left(\vartheta \theta_0 - 1\right)\dot{\gamma}_{\rm tar}^{*2}\tau^2
	\right]e^{-2\tau}.
\end{align}
The Mpemba effect takes place if  there exist positive real solutions $\tau$ of the quadratic equation
\begin{equation}
  F(\dot{\gamma}_{\rm tar}^*\tau)=0,
\end{equation}
where
\begin{equation}
	F(x)\equiv
	F_2 x^2 -\dot{\gamma}_{\rm ini}^* x + F_0,
\end{equation}
with
\begin{equation}
	F_2 \equiv \vartheta \theta_0-1,\quad
	F_0 \equiv 3 (\vartheta-1)\theta_0.
\end{equation}
Let us introduce the discriminant $D$ for $F(\tau)$ as
\begin{align}
	D
	&\equiv\dot{\gamma}_{\rm ini}^{*2} - 4F_0F_2\nonumber\\
	&= 6\left[-2\theta_0^2\vartheta^2 +2\theta_0(\theta_0+1)\vartheta-\theta_0 -1\right],
\end{align}
where we have used $\dot{\gamma}_{\rm ini}^{*2}=6(\theta_0-1)$. The two mathematical roots of $F(x)=0$ are
\begin{equation}
\label{Sol_x}
  x=\frac{\dot{\gamma}^*_{\rm ini}\pm\sqrt{D}}{2F_0}.
\end{equation}

Now we distinguish two cases:
\begin{enumerate}
  \item
  $\vartheta>1$.

  In this case, $F_0>0$ and $F_2>0$, so that the two  roots \eqref{Sol_x} are positive if $D>0$, while there is no real root if $D<0$. The latter possibility means that there is no Mpemba effect, but the former implies a ``NME+AME'' region. The boundary between both regions corresponds to $D=0$, which yields a quadratic equation for $\vartheta$  with a single solution with $\vartheta>1$.
  Therefore,
  \begin{equation}
  \label{NME+AME}
    1<\vartheta<\frac{12+\dot{\gamma}^{*2}_{\rm ini}+\dot{\gamma}^{*}_{\rm ini}\sqrt{12+\dot{\gamma}^{*2}_{\rm ini}}}{12+2\dot{\gamma}^{*2}_{\rm ini}}\Rightarrow\text{NME+AME}.
  \end{equation}
\item
 $\vartheta<1$.

 Now $F_0<0$. If, additionally, $F_2<0$, then no positive root of Eq.\ \eqref{Sol_x} is possible. However, if $F_2>0$, a single positive root exists (provided that $D>0$).
Therefore,
  \begin{equation}
  \label{AME}
    \left(1+\frac{1}{6}\dot{\gamma}^{*2}_{\rm ini}\right)^{-1}<\vartheta<1\Rightarrow\text{AME}.
  \end{equation}
  It can be easily checked that $D>0$ if Eq.\ \eqref{AME} is satisfied.
\end{enumerate}

\begin{figure}[htbp]
	\includegraphics[width=0.5\linewidth]{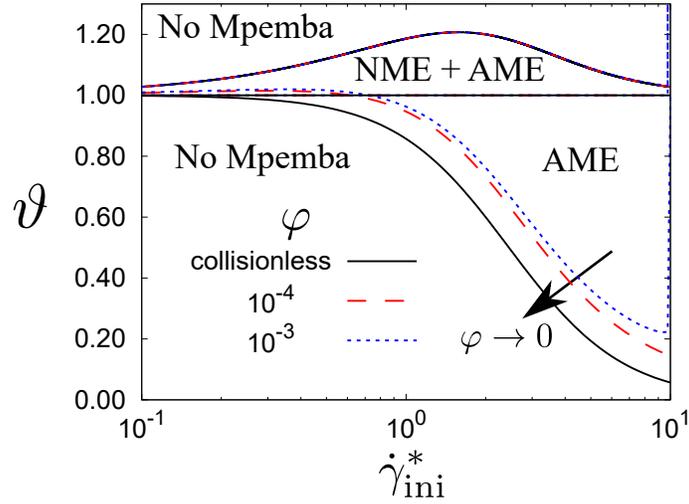}
	\caption{Phase diagram for the collisionless model in the cooling process.
	We also plot the phase diagrams for the collisional model for $\varphi=10^{-4}$ and $10^{-3}$ with  $e=0.9$, $\dot\gamma_{\rm tar}^*=1.0$, and $T_{\rm env}^{(\rm tar)*}=1.0$.
	}
	\label{fig:collisionless}
\end{figure}

Figure \ref{fig:collisionless} shows the phase diagram of the collisionless model in the cooling process, as described by Eqs.\ \eqref{NME+AME} and \eqref{AME}.
It is noteworthy that  the diagram is independent of the target shear rate $\dot{\gamma}^{*}_{\rm tar}$, which only determines the relaxation rate of the system.
For $\vartheta> 1$, a NME followed by a subsequent AME is observed with a maximum range $1<\vartheta<(1+\sqrt{2})/2\simeq 1.21$ at $\dot{\gamma}_{\rm ini}^*=\sqrt{6(\sqrt{2}-1)}\simeq 1.58$ (or, equivalently, $\theta_0=\sqrt{2}$).
The AME takes place easily for an increasing range of $\vartheta$ as the initial shear rate increases.
Note that the regions in Fig.\ \ref{fig:collisionless} where the Mpemba effect exists actually correspond to the inverse effect if $\dot{\gamma}^*_{\rm ini}<\dot{\gamma}^*_{\rm tar}$.

We also plot the phase diagrams for the collisional model for $\varphi=10^{-4}$ and $10^{-3}$ with the choice $e=0.9$, $\dot\gamma_{\rm tar}^*=1.0$, and $T_{\rm env}^{(\rm tar)*}=1.0$ in Fig.\ \ref{fig:collisionless}.
These results show that the behavior of the collisional model converges to that of the collisionless behavior in the low-density limit.

Similarly, let us now consider the evolution of the viscosity difference.
From Eq.\ \eqref{Pi_appA}, we obtain
\begin{equation}
	\Pi_{xy}^{{\rm FQE}*}(\tau) - \Pi_{xy}^{{\rm FS}*}(\tau)
	= \left[\frac{1}{2}\dot{\gamma}_{\rm ini}^*
		-\left(\vartheta\theta_0-1\right)\dot{\gamma}_{\rm tar}^* \tau\right]e^{-2\tau},
\end{equation}
or, equivalently,
\begin{equation}
	\eta_{\rm FQE}^*(\tau) - \eta_{\rm FS}^*(\tau)
	= -\left[\frac{1}{2}\frac{\dot{\gamma}_{\rm ini}^*}{\dot{\gamma}_{\rm tar}^*}
		-\left(\vartheta\theta_0-1\right) \tau\right]e^{-2\tau}.
	\label{eq:evol_eta_diff}
\end{equation}
This means that a (single) viscosity crossover takes place if $\vartheta\theta_0-1>0$, i.e., $\vartheta>1/({1+\frac{1}{6}\dot{\gamma}^{*2}_{\rm ini}})$. This includes the AME region \eqref{AME} plus the full region $\vartheta>1$. Thus, even if a Mpemba effect with $\vartheta>1$ is absent because Eq.\ \eqref{NME+AME} is not fulfilled, the difference $\eta_{\rm FQE}^*(\tau) - \eta_{\rm FS}^*(\tau)$ changes from negative to positive at a certain time.

\section{Relationship between the input parameters $\{T_{\rm env}^{(\rm ini)*}, T_{\rm env}^{({\rm tar}*)}, \dot{\gamma}_{\rm ini}^*\}$ and the outcome $\vartheta$ at $t=0$ }\label{sec:Langevin}
In this Appendix, we analyze the relationship between $\vartheta=T_{\rm FQE}(0)/T_{\rm FS}(0)$ and the input parameters $T_{\rm env}^{(\rm ini)*}$, $T_{\rm env}^{({\rm tar})*}$, and $\dot{\gamma}_{\rm ini}^*$.

\subsection{Equation for $T_{\rm FQE}(0)$}
\label{sec:B1}
In this subsection, we derive an approximate equation for the initial temperature $T_{\rm FQE}(0)$ in the FQE system as a function of $T_{\rm env}^{(\rm ini)}$. Since $T_{\rm FQE}(0)$ corresponds to the steady-state temperature for an unsheared system, we will  use in this subsection the notation
$T_{\rm FQE}(0)\to \theta_{\rm ini}T_{\rm env}^{(\rm ini)}$  for the sake of simplicity.
Note that  $\theta_{\rm ini}=1$ and $f(\bm{V})=f_{\text{eq}}(\bm{V})$ in the case of elastic collisions ($e=1$). However, if $e<1$, one has $\theta_{\rm ini}<1$ and $f(\bm{V})\neq f_{\text{eq}}(\bm{V})$.

Let us rewrite the unsheared  steady Boltzmann-Enskog equation \eqref{Enskog} in the dimensionless form
\begin{equation}
	\frac{\partial}{\partial \bm{c}}\cdot \left[\left(\bm{c} + \frac{1}{2\theta_{\rm ini}}\frac{\partial}{\partial \bm{c}}\right)\tilde{f}(\bm{c})\right]+ \frac{6}{\pi}\varphi \sqrt{2\theta_{\rm ini}T_{\rm env}^{(\rm ini)*}}\widetilde{J}_{\rm E}[\bm{c}| \tilde{f}, \tilde{f}]=0,
	\label{eq:dimensionless_Enskog}
\end{equation}
where we recall that $T_{\rm env}^*= T_{\rm env}/(m\sigma^2\zeta^2)$ and we have introduced the dimensionless velocity
\begin{equation}
	\bm{c}\equiv \frac{\bm{v}}{v_{T}}, \quad v_{T} \equiv \sqrt{\frac{2\theta_{\rm ini}T_{\rm env}^{(\rm ini)}}{m}}.
\end{equation}
In addition, we have also introduced the dimensionless distribution function $\tilde{f}(\bm{c})$ and collision integral $\widetilde{J}_{\rm E}[\bm{c}| \tilde{f}, \tilde{f}]$ as
\begin{subequations}
\begin{align}
	\label{eq:ftilde}
	\tilde{f}(\bm{c}) &\equiv \frac{v_{T}^3}{n}f(\bm{V})
	= \pi^{-3/2} e^{-c^2}\left[1+a_2 \left(\frac{c^4}{2}-\frac{5c^2}{2}+\frac{15}{8}\right)\right],\\
	\label{eq:J_E}
	\widetilde{J}_{\rm E}[\bm{c}| \tilde{f}, \tilde{f}] &\equiv \frac{v_{T}^2}{n^2 \sigma^2}J_{\rm E}[\bm{V}|f,f]
	= g_0 \int d\bm{c}_{2}\int d\widehat{\boldsymbol{\sigma}}\,
	\Theta (\widehat{{\boldsymbol {\sigma}}} \cdot \bm{c}_{12})
	(\widehat{\boldsymbol {\sigma }}\cdot \bm{c}_{12})
	\left[\frac{\tilde{f}(\bm{c}_1^{\prime\prime})\tilde{f}(\bm{c}_2^{\prime\prime})}{e^2}
	-\tilde{f}(\bm{c}_1)\tilde{f}(\bm{c}_2)\right].
\end{align}
\end{subequations}
In Eq.\ \eqref{eq:ftilde}, the distribution function has been expanded in Sonine polynomials and the expansion has been truncated after the second-order term, the coefficient $a_2$ representing the fourth velocity cumulant (or excess kurtosis).
It should be noted that the Enskog collision operator reduces to the Boltzmann collision operator multiplied by the radial distribution function at contact when the density and the temperature are uniform and the mean flow velocity vanishes (see, for instance, Appendix B in Ref.\ \cite{Hayakawa17}).
We also note that Eq.\ \eqref{eq:J_E} is obtained by taking the weak shear limit in Eq.\ \eqref{Enskog}, in which we ignore the finite core size effect in the collision integral.
As far as we have investigated, however, the error caused by this treatment is invisible, which will be shown later in this Appendix.

From Eq.\ \eqref{eq:dimensionless_Enskog}, and by neglecting terms nonlinear in $a_2$, one can obtain  \cite{Chamorro12, Garzo13PRE, Chamorro13}
\begin{subequations}
\label{eq:a2&T}
\begin{equation}
\theta_{\rm ini} =1-\frac{4}{\sqrt{\pi}}(1-e^2)\varphi g_0 \sqrt{T_{\rm env}^{(\rm ini)*}}\theta_{\rm ini}^{3/2}\left(1+\frac{3}{16}a_2\right),\label{eq:heat_eq_no_shear}
\end{equation}
\begin{equation}
	a_2=\frac{16(1-e)(1-2e^2)}{81-17e+30(1-e)e^2 + \displaystyle \frac{40\sqrt{\pi}}{(1+e)\varphi g_0 \theta_{\rm ini}^{3/2}\sqrt{T_{\rm env}^{(\rm ini)*}}}}.
\end{equation}
\end{subequations}
It is remarkable that $a_2\to 0$ and $\theta_{\rm ini}\to 1$ in the limit $\varphi\to 0$.
This suggests that the FQE is equivalent to a system at equilibrium in the low-density limit. In fact, Eq.\ \eqref{eq:dimensionless_Enskog} shows that collisions (either elastic or inelastic) become irrelevant in the limit $\varphi\to 0$. At finite density, on the other hand, $\theta_{\rm ini}$ can be obtained by numerically solving the set of coupled equations \eqref{eq:a2&T}. Once solved, we have $\theta_{\rm FQE}(0)=\theta_{\rm ini}T_{\rm env}^{(\rm ini)}/T_{\rm env}^{(\rm tar)}$.

\subsection{Explicit form of $\theta_{\rm ini}$ in the Maxwellian approximation}
\label{sec:B2}
In this subsection, we show the explicit form of $\theta_{\rm ini}$ by solving Eq.\ \eqref{eq:heat_eq_no_shear} when the excess kurtosis  $a_2$ is neglected.
In that approximation, Eq.\  \eqref{eq:heat_eq_no_shear} becomes
\begin{equation}
\label{eq:B5}
	\theta_{\rm ini}^{3/2}+\frac{\theta_{\rm ini}-1}{A}=0,\quad A\equiv\frac{4}{\sqrt{\pi}}(1-e^2)\varphi g_0 \sqrt{T_{\rm env}^{(\rm ini)*}}.
\end{equation}
Introducing the change of variable $\theta_{\rm ini}=[x-1/(3A)]^2$,  we can rewrite Eq.\ \eqref{eq:B5} as
\begin{equation}
	x^3 + C_1 x+ C_0=0, \quad C_1 \equiv -\frac{1}{3A^2},\quad	C_0 \equiv -\frac{1}{A} + \frac{2}{27A^3}.
\label{eq:theta1_eq}
\end{equation}

We  define the discriminant of the cubic equation \eqref{eq:theta1_eq} as
\begin{equation}
	\Delta = \left(\frac{C_0}{2}\right)^2+\left(\frac{C_1}{3}\right)^3
	=\frac{1}{4A^2}\left(1-\frac{4}{27A^2}\right).
\end{equation}
If $\Delta\ge 0$ (i.e., $A\geq 2/3\sqrt{3}\simeq 0.385$), the cubic equation \eqref{eq:theta1_eq} has only one real root given by
\begin{equation}
	x =\ \sqrt[3]{\frac{1}{2A}-\frac{1}{27A^3}+\frac{1}{2A}\sqrt{1-\frac{4}{27A^2}}}
	+ \sqrt[3]{\frac{1}{2A}-\frac{1}{27A^3}-\frac{1}{2A}\sqrt{1-\frac{4}{27A^2}}}.
\end{equation}
On the other hand, if $\Delta< 0$ (i.e., $A< 2/3\sqrt{3}\simeq 0.385$), the cubic equation \eqref{eq:theta1_eq} has three real roots, one positive and two negative. It can be checked that only the positive root is consistent with the physical condition $\theta_{\rm ini}<1$. After using Vieta's method, we obtain
\begin{equation}
	x = \frac{2}{3A}\cos\left[\frac{1}{3}\cos^{-1}\left(\frac{27}{2}A^2-1\right)\right].
\end{equation}

In summary, within the Maxwellian approximation, we have
\begin{equation}
\label{eq:thetaini}
	\theta_{\rm ini}
	=
	\begin{cases}
		\displaystyle \frac{1}{9A^2}\left(\sqrt[3]{\frac{27}{2}A^2-1+\frac{27}{2}A^2\sqrt{1-\frac{4}{27A^2}}}
		+ \sqrt[3]{\frac{27}{2}A^2-1-\frac{27}{2}A^2\sqrt{1-\frac{4}{27A^2}}}-1\right)^2,& \displaystyle{A\geq \frac{2}{3\sqrt{3}}},\\
		\displaystyle \frac{1}{9A^2} \left\{2\cos\left[\frac{1}{3}\cos^{-1}\left(\frac{27}{2}A^2-1\right)\right]-1\right\}^2,& \displaystyle{A< \frac{2}{3\sqrt{3}}}.
	\end{cases}
\end{equation}
It can be checked that the values of $\theta_{\rm ini}$ obtained from the numerical solution of Eqs.\ \eqref{eq:a2&T} are practically indistinguishable from those given by Eq.\ \eqref{eq:thetaini}.

\begin{figure}[htbp]
	\includegraphics[width=0.9\linewidth]{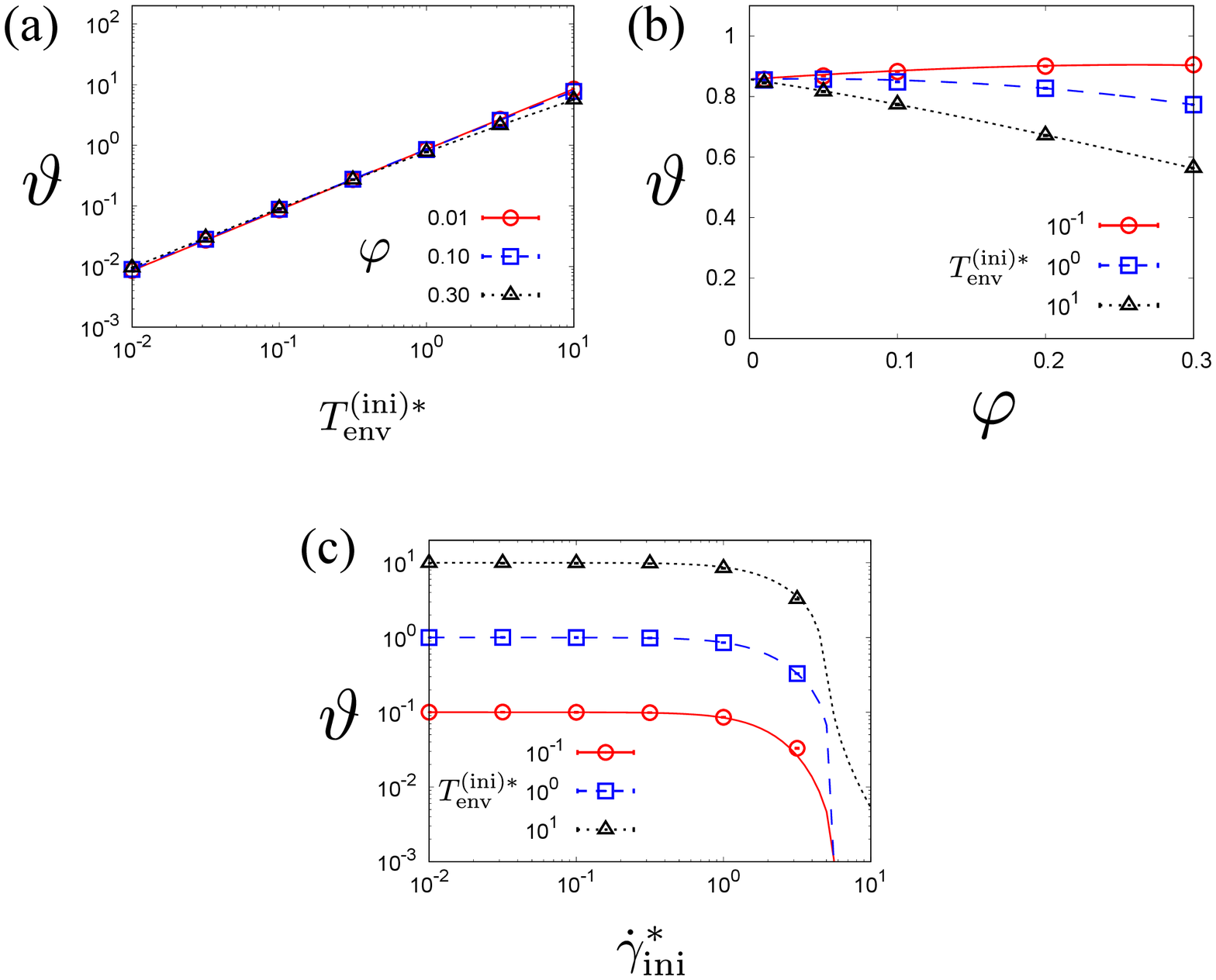}
	\caption{(a) Plots of $\vartheta$ against $T_{\rm env}^{(\rm ini)*}$ for $\varphi=0.01$, $0.10$, and $0.30$ with fixed $e=0.9$, $\dot\gamma_{\rm ini}^*=1.0$, and $T_{\rm env}^{(\rm tar)*}=1.0$.
	(b) Plots of $\vartheta$ against $\varphi$ for $T_{\rm env}^{(\rm ini)*}=0.10$, $1.0$, and $10$ with fixed $e=0.9$, $\dot\gamma_{\rm ini}^*=1.0$, and $T_{\rm env}^{(\rm tar)*}=1.0$.
	(c) Plots of $\vartheta$ against $\dot\gamma_{\rm ini}^*$ for $T_{\rm env}^{(\rm ini)*}=0.10$, $1.0$, and $10.0$ with fixed $e=0.9$, $\varphi=0.01$, and $T_{\rm env}^{(\rm tar)*}=1.0$.
	The symbols (with error bars) are obtained from our simulations and the lines are kinetic-theory predictions.}
	\label{fig:theta_eq}
\end{figure}

\subsection{Relationship between the input parameters and the outcome}
In Secs.\ \ref{sec:B1} and \ref{sec:B2}, we have obtained  $\theta_{\rm FQE}(0)=\theta_{\rm ini} T_{\rm env}^{(\rm ini)}/T_{\rm env}^{(\rm tar)}$.
On the other hand, $\theta_{\rm FS}(0)$ is numerically determined from the set of Eqs.\ \eqref{eq:evol_theta-eq:evol_Pi_xy} by setting $\partial_\tau\to 0$,  $\dot\gamma^*\to \dot\gamma_{\rm ini}^*$, and $T_{\rm env}^{*}\to T_{\rm env}^{(\rm tar)*}$. This in turn provides $\vartheta=\theta_{\rm FQE}(0)/\theta_{\rm FS}(0)$ as a function of the input parameters $T_{\rm env}^{(\rm ini)*}$, $T_{\rm env}^{(\rm tar)*}$, and $\dot\gamma_{\rm ini}^*$.

Figure \ref{fig:theta_eq} shows the dependencies of $\vartheta$ on (a) the environmental temperature $T_{\rm env}^{(\rm ini)*}$ in FQE, (b) the packing fraction $\varphi$, and (c) the initial shear rate $\dot\gamma_{\rm ini}^*$ in FS.
As can be seen, the theory reproduces  very well our simulation results.
As $T_{\rm env}^{(\rm ini)*}$ increases, the temperature ratio $\vartheta$ monotonically increases [see Fig.\ \ref{fig:theta_eq}(a)]. Also, Fig.\ \ref{fig:theta_eq}(b) shows that $\vartheta$ tends to decrease with increasing density.
Moreover, $\vartheta$ decreases as $\dot\gamma_{\rm ini}^*$ increases, as shown in Fig.\ \ref{fig:theta_eq}(c), but this effect is only remarkable for very large values of initial shear rate $\dot\gamma_{\rm ini}^*$.

\section{Crossing times of the temperatures at Mpemba effect}\label{sec:Mpemba_time}
\begin{figure}[htbp]
	\includegraphics[width=0.9\linewidth]{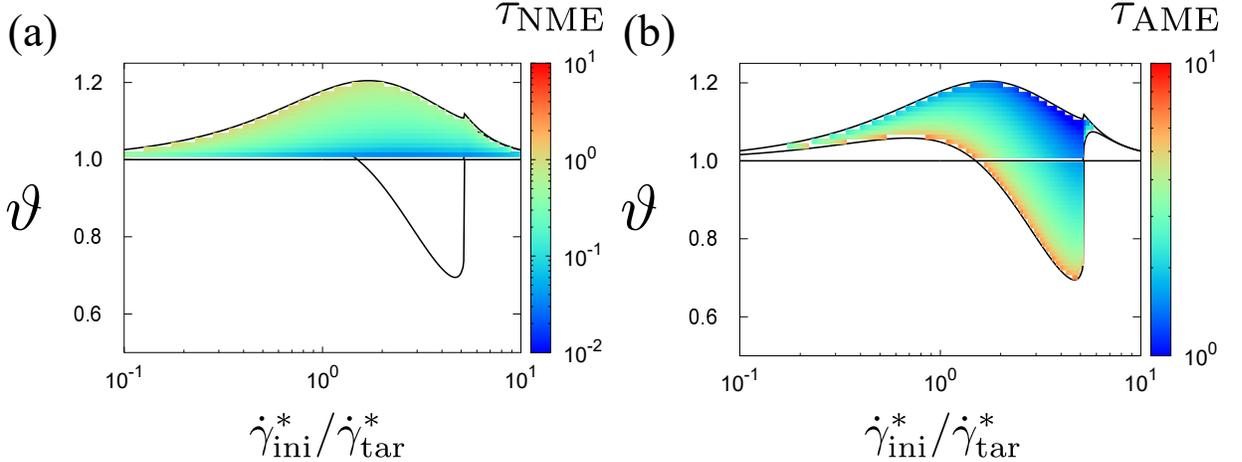}
	\caption{Magnitude plots of the times (a) $\tau_{\rm NME}$ and (b) $\tau_{\rm AME}$ for $\varphi=0.01$, $e=0.9$, $T_{\rm env}^{(\rm tar)*}=1.0$, and $\dot\gamma_{\rm tar}^*=1.0$.
	These phase diagrams are identical to Fig.\ \ref{fig:phase_xi}(a).}
	\label{fig:Mpemba_time}
\end{figure}

In this Appendix, we investigate the times $\tau_{\rm NME}$ and $\tau_{\rm AME}$, which are defined by Eqs.\ \eqref{eq:tau_NME_AME_def1} and \eqref{eq:tau_NME_AME_def2}.
Figure \ref{fig:Mpemba_time}(a) shows the magnitude plot of $\tau_{\rm NME}$ for $\varphi=0.01$, $e=0.9$, $T_{\rm env}^{(\rm tar)*}=1.0$, and $\dot\gamma_{\rm tar}^*=1.0$.
This time remains finite near the boundary between ``No Mpemba'' and ``NME+AME,'' and between ``NME'' and ``NME+AME,'' which is natural because NME takes place in the early stage of evolution.
On the other hand, this time tends to zero on the boundaries between ``NME'' and ``No Mpemba,'' and between ``NME+AME'' and ``AME'' because the initial temperature difference disappears on the boundaries ($\vartheta\gtrsim1$).

Figure \ref{fig:Mpemba_time}(b) also shows the magnitude plot of $\tau_{\rm AME}$ under the identical set of parameters to that used in Fig.\ \ref{fig:Mpemba_time}(a).
This time also remains finite near the boundary between ``No Mpemba'' and ``NME+AME'' (in fact $\tau_{\rm AME}$ and $\tau_{\rm NME}$ coalesce on that boundary), because the AME takes place due to the early stage of evolution.
On the other hand, $\tau_{\rm AME}$ diverges on the boundaries between ``NME+AME'' and ``NME,'' and between ``AME'' and ``No Mpemba'' because such crossings take place in the later stage of evolution as shown in Figs.\ \ref{fig:evol_xi}(d) and (f).

\section{Domain structures near the boundaries}\label{sec:domain}
\begin{figure}[htbp]
	\includegraphics[width=0.85\linewidth]{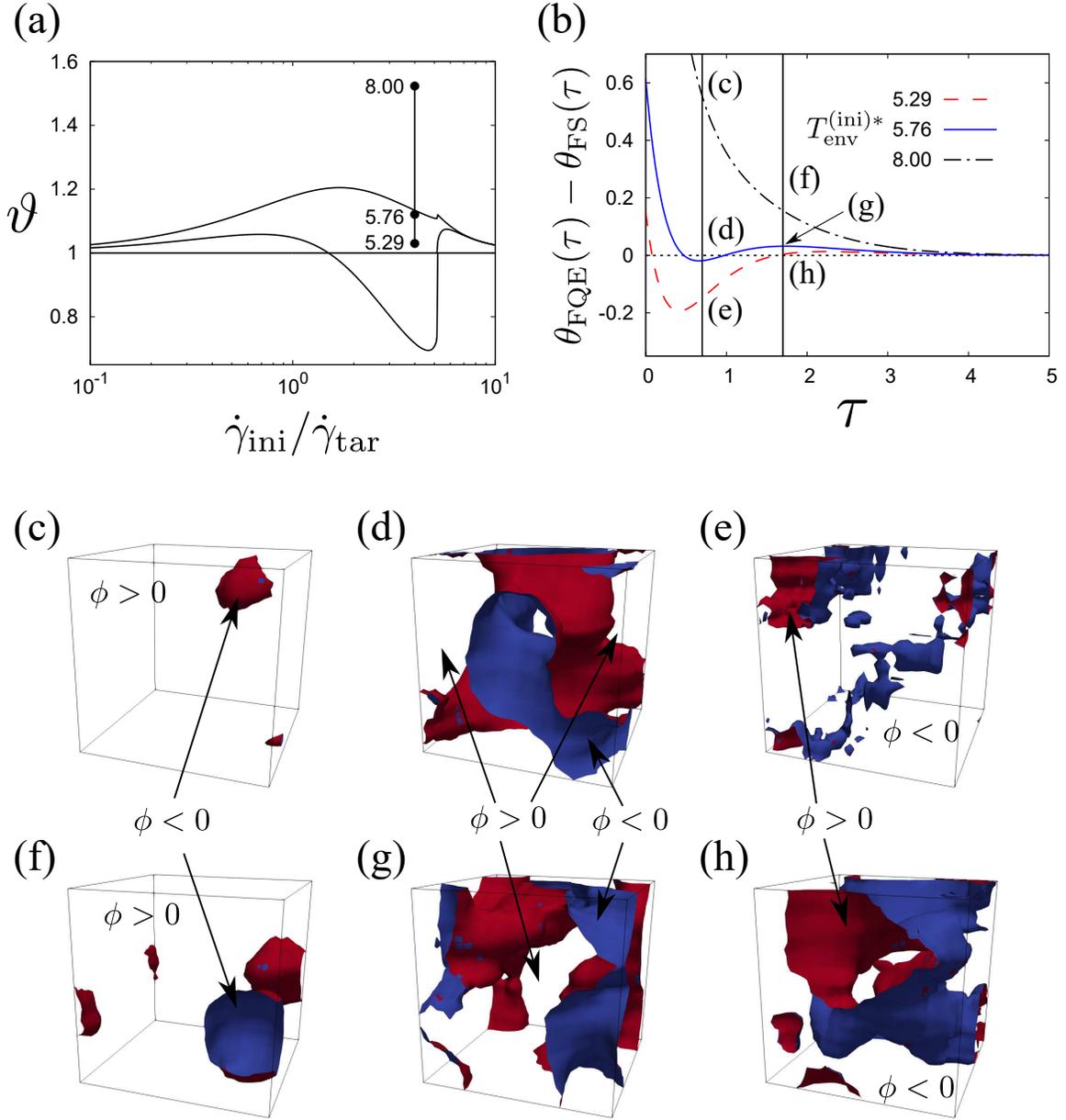}
	\caption{(a) Phase diagram in the plane $\vartheta$ versus $\dot\gamma_{\rm ini}/\dot\gamma_{\rm tar}$ of the Mpemba effect for $\varphi=0.01$, $e=0.9$, $T_{\rm env}^{({\rm tar})*}=1.0$, and $\dot\gamma_{\rm tar}^*=1.0$.
	We have chosen three initial environmental temperature $T_{\rm env}^{(\rm ini)*}=5.29$, $5.76$, and $8.00$.
	(b) Time evolutions of the temperature differences at $T_{\rm env}^{(\rm ini)*}=5.29$ (red dashed line), $5.76$ (blue solid line), and $8.00$ (black dotted line).
	Here, we have chosen six points to visualize the order parameter defined in Eq.\ \eqref{eq:def_phi}.
	Points (c)--(e) and (f)--(h) correspond to the temperature differences at $\tau=0.7$ and $1.7$, respectively.
	(c)--(e) Contours of $\phi(\bm{r}^*, \tau)=0$ at $\tau=0.7$ and $T_{\rm env}^{({\rm ini})*}=8.00$, $5.76$, and $5.29$, respectively, where the red (blue) side on the contours is higher (lower) than $0$.
	(f)--(h) Contours of $\phi(\bm{r}^*,\tau) = 0$ at $\tau=1.7$ in which the other parameters are identical those used in Figs.\ \ref{fig:domain}(c)--(f).
	The small minority phases correspond to $\phi<0$ ($\phi>0$) in panels (c) and (f) (panel (e)).}
	\label{fig:domain}
\end{figure}
In this Appendix, we discuss domain structures near the phase boundaries by the introduction of an order parameter.
To this end, we first evaluate the temperature fields $\theta_{\rm FQE}(\bm{r}^*, \tau)$ and $\theta_{\rm FS}(\bm{r}^*, \tau)$, where $\bm{r}^*\equiv \bm{r}/L$ is the dimensionless coordinate with the linear system size $L$, and the mesh size is chosen as $\Delta r^* = 1/30$.
We introduce the order parameter field
\begin{equation}
	\phi(\bm{r}^*,\tau) 
	\equiv \frac{\theta_{\rm FQE}(\bm{r}^*, \tau) - \theta_{\rm FS}(\bm{r}^*, \tau)}
		{\theta_{\rm FQE}(\bm{r}^*, \tau) +\theta_{\rm FS}(\bm{r}^*, \tau)}. \label{eq:def_phi}
\end{equation}
We note that this order parameter is defined in terms of the two different simulations.
Let us check how this order parameter behaves in the phase diagram.
We choose three points which belong to the ``No Mpemba,'' ``NME+AME,'' and close to the phase boundary between these two phases, as shown in Fig.\ \ref{fig:domain}(a).
Typical time evolutions of the temperature difference are plotted in Fig.\ \ref{fig:domain}(b).
Here, we choose two (dimensionless) times $\tau=0.7$ and $1.7$ which correspond to the times when the temperature differences near the phase boundary ($T_{\rm env}^{(\rm ini)*}=5.76$) take extrema.
Figures \ref{fig:domain}(c)--(e) and (f)--(h) illustrate the contours of $\phi(\bm{r}^*, \tau)=0$ at $\tau=0.7$ and $1.7$, respectively.
At $T_{\rm env}^{({\rm ini})*}=8.00$, the system belongs to ``No Mpemba'' and far from the boundary.
In this case, the majority phase in the order parameter field takes almost positive values ($\phi>0$) as can be seen in Figs.\ \ref{fig:domain}(c) and (f).
This means that small ``NME'' domains can exist but they cannot dominate the space.
Similar behavior can be also observed at $T_{\rm env}^{(\rm ini)*}=5.29$ and $\tau=0.7$ (see Fig.\ \ref{fig:domain}(e)) in which small ``No Mpemba'' domains ($\phi<0$) exist in the background of majority ``NME'' ($\phi>0$) phase.
On the other hand, we observe bicontinuous domain structures at $T_{\rm env}^{(\rm ini)*}=5.76$.
A small ``NME'' phase starts to increase its size in the early stage.
When the temperature difference take extrema, the domains become bicontinuous at $\tau=0.7$ and $1.7$ (see Figs.\ \ref{fig:domain}(d) and (g), respectively).
This means that the ``NME'' (``AME'') phase competes with the ``No Mpemba'' (``NME'') at $\tau=0.7$ (1.7).
Here, the structure is connected because of the adoption of the periodic boundaries.
We note that point (h) of Fig.\ \ref{fig:domain}(b) also exhibits bicontinous domains at which the temperature difference becomes zero.
The corresponding domain structure should be observed on each phase boundary, though we do not draw such figures explicitly in this paper.

\section{Evolution of the viscosity difference}\label{sec:eta}

\begin{figure}[htbp]
	\includegraphics[width=0.9\linewidth]{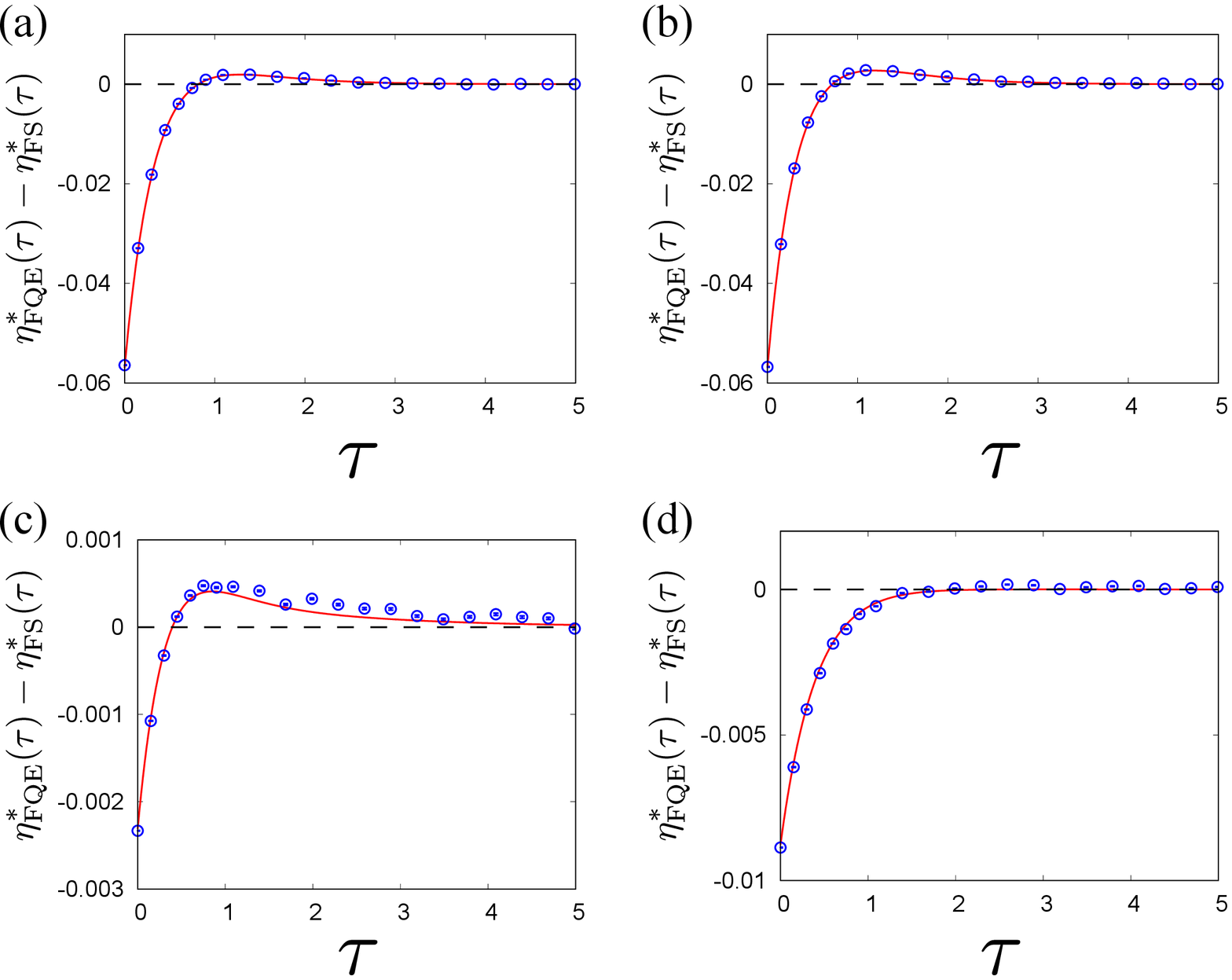}
	\caption{Time evolutions of the viscosity difference $\eta_{\rm FQE}^*(\tau)-\eta_{\rm FS}^*(\tau)$ for $e=0.9$,  $\varphi=0.01$, $T_{\rm env}^{(\rm tar)*}=1.0$, and (a)  $\dot{\gamma}_{\rm tar}^*=1.0$, $\dot{\gamma}_{\rm ini}^*=4.0$, and $T_{\rm env}^{(\rm ini)*}=5.29$, (b) $\dot{\gamma}_{\rm tar}^*=1.0$, $\dot{\gamma}_{\rm ini}^*=4.0$, and $T_{\rm env}^{(\rm ini)*}=5.76$, (c) $\dot{\gamma}_{\rm tar}^*=4.0$, $\dot{\gamma}_{\rm ini}^*=1.0$, and $T_{\rm env}^{(\rm ini)*}=1.33$, and (d) $\dot{\gamma}_{\rm tar}^*=1.0$, $\dot{\gamma}_{\rm ini}^*=0.95$, and $T_{\rm env}^{(\rm ini)*}=1.21$
	The parameters used in panels (a), (b), (c), and (d) correspond to those in Figs.\ \ref{fig:evol_xi}(a), \ref{fig:evol_xi}(c), \ref{fig:evol_xi_inverse}, and \ref{fig:evol_mixed}, respectively.
	The symbols (with error bars) are obtained from our simulations and the lines are kinetic-theory
predictions.}
	\label{fig:eta}
\end{figure}

Since the viscosity is the most important quantity for the rheology of suspensions, the analysis of its transient behavior might be more relevant than that of temperature.
In this Appendix, we present some typical results for the time evolution of the viscosity difference between two systems which exhibit Mpemba effect.

Figure \ref{fig:eta} shows the evolutions of the viscosity difference $\eta_{\rm FQE}^*(\tau)-\eta_{\rm FS}^*(\tau)$  for the cases considered in Figs.\ \ref{fig:evol_xi}(a), \ref{fig:evol_xi}(c), \ref{fig:evol_xi_inverse}, and \ref{fig:evol_mixed}.
It is notable that the viscosity difference changes its sign only once, as expected from Eq.\ \eqref{eq:evol_eta_diff} in the collisionless model. 
This is a feature of the viscosity different from that of the Mpemba effect for the temperature relaxation.


\end{document}